\documentclass[a4paper]{article}
\usepackage[text={7in,10in},centering]{geometry}
\usepackage{graphicx,color}
\usepackage{amsmath,amsthm,amssymb}
\usepackage{multirow,array}
\usepackage[table]{xcolor}
\usepackage{longtable}
\usepackage{caption,subcaption}
\usepackage{url}
\usepackage{algorithmic} 
\usepackage[ruled,vlined]{algorithm2e}
\usepackage{hyperref}
\usepackage{cleveref}

\newcommand{\ci}{\mathrm{CI}} 
\newcommand{\wm}{\mbox{W/m}^2} 
\newcommand{\kw}{\mbox{kW}} 
\newcommand{\kwhm}{\mbox{kWh/m}^2} 
\newcommand{\sqkm}{\mbox{km}^2}
\newcommand{\iclr}{I_{\mathrm{clr}}}
\newcommand{\am}{\mathrm{AM}}
\newcommand{\lat}{\mbox{lat}}
\newcommand{\llong}{\mbox{long}}
\newcommand{\hour}{\mbox{hour}} 
\newcommand{\dday}{\mbox{day}} 
\newcommand{\mmonth}{\mbox{month}}
\newcommand{\degreeNorth}{$^{\circ}\text{N}$} 
\newcommand{\degreeEast}{$^{\circ}\text{E}$} 

\newcommand{\Real}{\mathbf{R}} 
\newcommand{\FeatureSeq}{X} 
\newcommand{\Model}{f} 
\newcommand{\Predicted}{\hat{I}} 
\newcommand{\GroundTruth}{I}

\newcommand{\seqlen}{p} 
\newcommand{\Set}[1]{\left\{ #1 \right\}} 
\newcommand{\Brc}[1]{\left( #1 \right)}
\newcommand{\RBrc}[1]{\left[ #1 \right]}

\newcommand{\softmax}{\text{Softmax}} 
\newcommand{\selfAttention}{\text{Atten}} 
\newcommand{\sparseselfAttention}{\text{SparseAtten}}  

\newcommand\ie{\textit{i.e.}}
\newcommand\eg{\textit{e.g.}}
\newcommand\approxtext{\textit{approx.}}

\newcommand{\tablefontsize}{\footnotesize}

\usepackage{array}

\newcolumntype{L}[1]{>{\raggedright\let\newline\\\arraybackslash\hspace{0pt}}m{#1}}
\newcolumntype{C}[1]{>{\centering\let\newline\\\arraybackslash\hspace{0pt}}m{#1}}
\newcolumntype{R}[1]{>{\raggedleft\let\newline\\\arraybackslash\hspace{0pt}}m{#1}}

\definecolor{buff}{rgb}{0.94, 0.86, 0.51}
\definecolor{dandelion}{rgb}{0.94, 0.88, 0.19}
\definecolor{lightapricot}{rgb}{0.99, 0.84, 0.69}
\definecolor{lightcoral}{rgb}{0.94, 0.5, 0.5}
\definecolor{mediumspringbud}{rgb}{0.79, 0.86, 0.54}
\definecolor{palecornflowerblue}{rgb}{0.67, 0.8, 0.94}
\definecolor{pink-orange}{rgb}{1.0, 0.6, 0.4}
\definecolor{lightgray}{rgb}{0.83, 0.83, 0.83}

\graphicspath{{figures/}}

\title{Developing a Thailand solar irradiance map using Himawari-8 satellite imageries and deep learning models}

\author{Suwichaya Suwanwimolkul, Natanon Tongamrak,Nuttamon Thungka \\
 Naebboon Hoonchareon, and Jitkomut Songsiri\footnote{Corresponding author} \\[1ex]
 Department of Electrical Engineering, Faculty of Engineering \\ [1ex]
 Chulalongkorn University, Bangkok, Thailand 10330 \\[1ex]
 e-mail: suwichaya.s@chula.ac.th, natanon.t@chula.ac.th, nuttamon.thungka@gmail.com \\ naebboon.h@chula.ac.th, and jitkomut.s@chula.ac.th
}   

\begin{document}
\maketitle

\begin{abstract}
This paper presents an online platform showing Thailand solar irradiance map every 30 minutes available at \url{https://www.cusolarforecast.com}. The methodology for estimating global horizontal irradiance (GHI) across Thailand relies on cloud index extracted from Himawari-8 satellite imagery, Ineichen clear-sky model with locally-tuned Linke turbidity, and machine learning models. The methods take clear-sky irradiance, cloud index, re-analyzed GHI and temperature data from the MERRA-2 database, and date-time as inputs for GHI estimation models, including LightGBM, LSTM, Informer, and Transformer. These are benchmarked with the estimate from a commercial service X by evaluation of 15-minute ground GHI data from 53 ground stations over 1.5 years during 2022-2023. The results show that the four models exhibits comparable overall MAE performance to the service X. The best model is LightGBM with an overall MAE of 78.58 $\wm$ and RMSE of 118.97 $\wm$, while the service X achieves the lowest MAE, RMSE, and MBE in cloudy condition. Obtaining re-analyzed MERRA-2 data for the whole Thailand region is not economically feasible for deployment. When removing these features, the Informer model has a winning performance in MAE of 78.67 $\wm$. The obtained performance aligns with existing literature by taking the climate zone and time granularity of data into consideration. As the map shows an estimate of GHI over 93,000 grids with a frequent update, the paper also describes a computational framework for displaying the entire map. It tests the runtime performance of deep learning models in the GHI estimation process.
\end{abstract}


\section{Introduction}
\label{sec:intro}

Thailand has targeted to achieve carbon neutrality by 2050 when the power grid will need to accommodate 50\% share of renewable electricity generation capacity; see \cite{eppo2022}. The most recent draft of Power Development Plan 2024 (PDP2024) for 2024 – 2037 from~\cite{PDP2024} proposes to add a new solar generation capacity of approximately 24,400 MWp (more than 4 times the amount issued in the previous Alternative Energy Development Plan 2015–2036 (AEDP2015) at 6,000 MWp, shown in~\cite[p.9]{eppo2015}. This amount does not yet include behind-the-meter, self-generation solar installed capacities of the prosumers, which is expected to increase at an accelerating rate. Solar integration into the power grid with such a sharp-rising amount will pose technical challenges to the operation and control of the transmission and distribution networks, carried out by the transmission system operator (TSO) and distribution system operator (DSO), as presented in~\cite{Obi2016}. Hence, TSO in Thailand will need an effective means to estimate the solar power generation across the entire transmission network, on an hourly basis, or even finer time resolution, to provide economic hour-to-hour generation dispatch for load following the total net load of the transmission, and to prepare sufficient system flexibility (\ie, ramp-rate capability of the thermal and hydropower plants, or energy storage systems) to cope with the net load fluctuation due to solar generation intermittency for maintaining system frequency stability, concurrently, in its operation. For DSO, a significant amount of reverse power flow when self-generation from solar exceeds self-consumption can lead to technical concerns of voltage regulation and equipment overloading problems. The near real-time estimation of solar generation in each distribution area will enable DSO to activate proper network switching or reconfiguring to mitigate such fundamental concerns to ensure its reliable operation.

While solar energy offers a clean and virtually infinite alternative to fossil fuels, its dependence on sunshine makes it susceptible to fluctuations in weather patterns. This variability, potentially exacerbated by climate change, necessitates the development of accurate solar forecasting technology. The reliability of these forecasts directly impacts the planning (pipeline) of solar photovoltaic (PV) systems, ensuring they are integrated efficiently into the power grid, and also influences the operational settings required for various energy applications.
 
Global horizontal irradiance (GHI) estimation services provide crucial information in the solar energy industry. These services often provide estimated GHI for a desired location by using a combination of weather models, satellite imagery, and measurements from ground stations. These GHI databases can be divided into i) satellite-derived databases where GHI data were estimated from satellite imagery products, ii) ground-based measurements containing GHI recorded from available monitoring stations, and iii) reanalysis datasets that fuse weather estimates from atmospheric models with historical observations, including GHI data. According to \cite{Polo2016}, \Cref{tab:solarservice} in the Appendix lists some satellite-derived solar irradiance services that differ from the methods they use and service coverage, dependent on the satellite. Some of the leading services that provide estimates of solar irradiance and power forecasts with global coverage can be described as follows. 

SolarAnywhere provides a core technique called the SUNY model developed in the works from \cite{PerezSolarAnywhere2013, Perez2002, Perez2015, Perez2018} that uses satellite imagery and numerical weather predictions to better calculate irradiance under cloudy conditions. SolarGIS by \cite{SuriSolarGIS2010} uses a semi-empirical solar radiation model with data from satellites for explaining cloud properties. The service also provides site adaptation of SolarGIS data to improve the quality using ground measurements. PVWatts by NREL;see \cite{PVWatts2013}, has an interactive page where the user can input a desired location and PV panel's parameters to calculate monthly solar irradiance and PV power. \cite{sodapro} offers solar radiation maps in 15-min resolution. The method relies on HelioClim-3 persistence forecast of a clear-sky index and the use of the Meteosat satellite. SolCast service, detailed in \url{https://www.solcast.com/irradiance-data-methodology}, provides solar irradiance forecasts and historical data in several time resolutions. It uses NOAA, EUMETSAT which covers the continental US, Europe, and Himawari-8 satellite from JMA, which covers Asia and the Pacific, to gather information about cloud opacity and weather conditions. The methodology of producing GHI estimates consists of the SolCast cloud model and the REST2v5 clear-sky model. The SolCast cloud model takes historical and forecast data of geostationary meteorological satellite (GMS) and NWP inputs for 14 days ahead. The cloud opacity extracted from satellite cloud detection algorithms is combined with NWP inputs to account for mixing effects from weather artifacts. The validation of Solcast was presented in \url{https://www.solcast.com/validation-and-accuracy} for all climate conditions where sites are grouped by continent. The overall normalized MAE and RMSE of SolCast 2023 based on all 207 sites were $10.33\%$ and $15.99\%$, respectively. The service also provides the SolCast Rooftop PV model, which estimates PV power when users provide PV metadata such as location, capacity, and panel tilt angle.


\begin{figure}[ht]
\centering
\begin{subfigure}[b]{0.45\linewidth}
\centering
\includegraphics[height = 5cm]{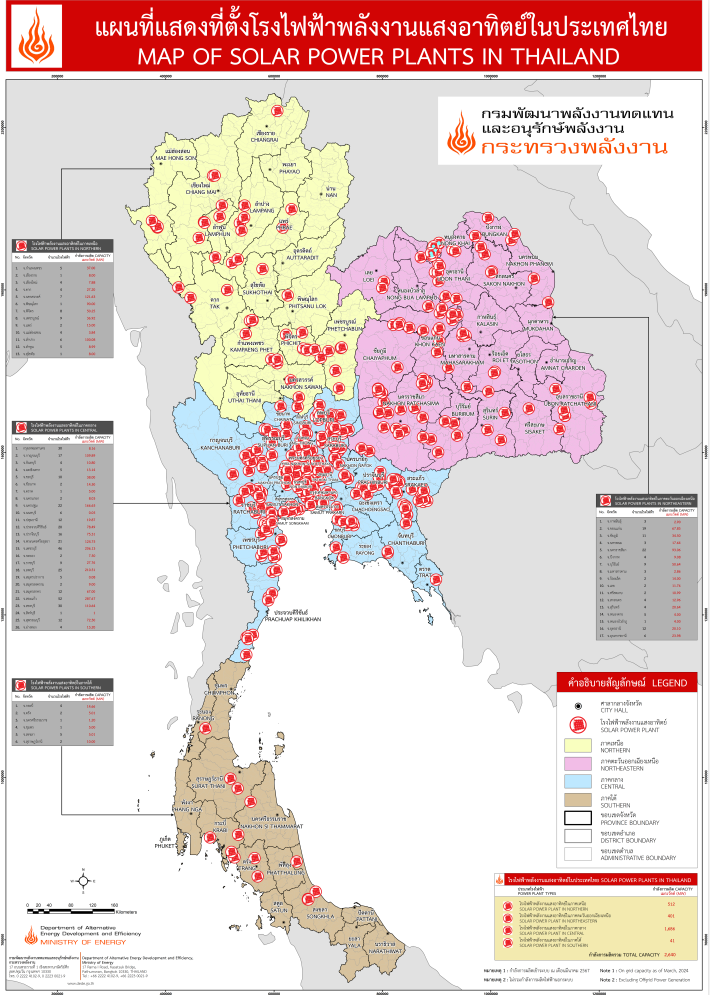}
\caption{The total on-grid solar plant capacity is 2,640 MW.} 
\end{subfigure}
\begin{subfigure}[b]{0.45\linewidth}
\centering
\includegraphics[height = 5cm]{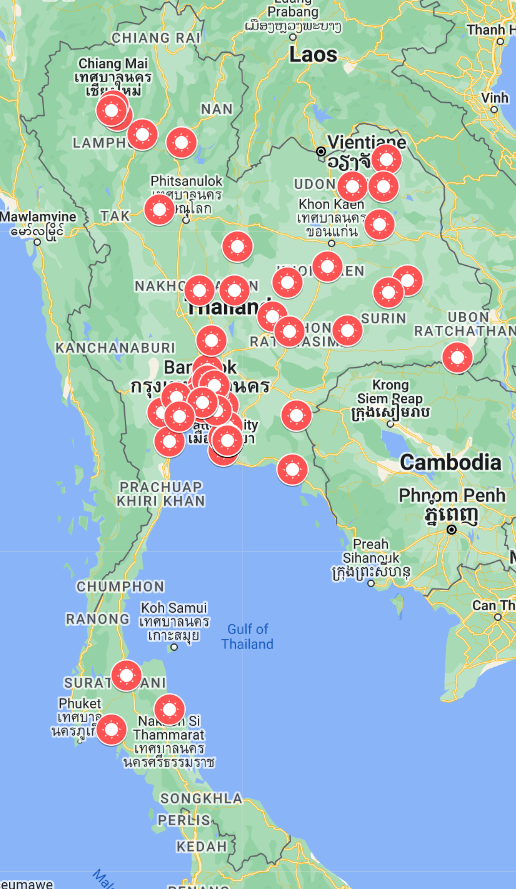}
\caption{53 solar plants with ground measurements.} 
\end{subfigure}
\caption{Solar power plant data in Thailand (as of March 2024). Photo credit: Renewable energy and conservation GIS of Thailand \url{https://gis.dede.go.th/}.}
\label{fig:solarsite}
\end{figure}

While these aforementioned companies provide a wide range of services and use advanced algorithms with several sources of input data, it boils down to the cost of acquiring these solar data for a regional application as the cost often varies with the number of sites. As of March 2024, Thailand has 552 on-grid solar power plants with a total capacity of 2,640 MW, shown in \Cref{fig:solarsite} (a). Among these solar sites, \Cref{fig:solarsite} (b) shows 53 solar sites fully equipped with ground measurements that participate in our study. To accommodate solar energy applications in Thailand in an operational setting, we aim to develop our solar map that frequently estimates satellite-derived GHI across the country using satellite imagery and local ground measurements. The cost of acquiring irradiance estimates across Thailand (93,061 grids) through commercial services in an operational setting (every 30 minutes) is high as compared to the cost of developing our own platform. For this reason, we study the literature on satellite-derived GHI estimation techniques in the following sections.

\subsection{Physic-based and empirical methods} 
\label{sec:empirical}
Traditionally, physical models use satellite information to solve the radiative transfer model, which simulates how radiation moves through absorption and scatter processes in the Earth's atmosphere. Empirical techniques rely on the statistical relationship between atmospheric parameters and ground observations. The two approaches have been merged into semi-empirical models, which analytically describe a relationship between solar radiation and aerosols in the atmosphere while using satellite information to empirically retrieve the cloud index. 

Methods in this group quantify how much clouds and other particles in the atmosphere affect the incoming solar radiation by compensating with a clear-sky scenario. This relationship is often parametrized in physical parameters such as aerosol optical depth (AOD), reflectance, surface albedo, which can be derived from satellite imagery. The Heliosat model is a semi-empirical method with a core principle that combines a clear-sky model with a cloud index through a piecewise-polynomial relationship. \cite{Dagestad2005} used Heliosat algorithm and calculated a new cloud index from reflectivity derived from the Meteosat satellite where a modification was made on computing the ground reflectivity to save computing time. \cite{Kashyap2018} extracted the brightness temperature difference from various bands of INSAT-3-D satellite to calculate the all-day index as information of cloud, dust, and fog. The GHI estimation relied on the Heliosat-2 model and was validated on ground measurements in India. The RMSE ranged from 48 to 128 $\wm$ under sky conditions with dust and fog. \cite{Valor2023} presented an improvement over the MAGIC-Heliosat method on calculating apparent reflectance to compute the cloud index. The results were validated with ground stations from the baseline surface radiation network (BSRN) in Australia and Japan. The method's performance varied according to evaluation on hourly, daily, and monthly averaged data with RMSE of 161, 80, and 70 $\wm$, respectively. 

While the Heliosat model is a well-known method, there are several other semi-empirical approaches for GHI estimation. \cite{Qin2018} showed that Yang's hybrid model (YHM) performed best with RMSE of 27.94 $\wm$ when compared to ANN on datasets in China. The physical method YHM relies on the radiation-damping processes, \eg, Rayleigh scattering, aerosol extinction, absorption by ozone, water vaport and permanent gas. These atmospheric variables were derived from MODIS satellite. \cite{Qin2019} presented a mathematical formula of GHI as a function of irradiance with zero surface albedo, global reflectance function of atmosphere and surface albebo, which can be derived from Himawari-8 satellite. The results were validated on 3-year ground station data in Australia with MAE of 55.8 $\wm$ and RMSE of 91.7 $\wm$. The GHI model with explicit surface albedo was later used in \cite{Qin2021} but all irradiance components were calculated by pre-computing Look Up Table and using radiative transfer model (RTM). The GHI estimation results were validated on data in Australia with an MAE of 54.7 $\wm$ and an RMSE of 90.3 $\wm$. \cite{Bright2021} computed the cloud indices using the surface albedo extracted from eight-band imagery of Himawari-8. These cloud indices were converted to clear-sky indices (for all eight bands), and the eight-band clear-sky indices were optimally weighted. The results were validated using data in Singapore, where the scatter plot between the ground-observed clear-sky index and the satellite-estimated values was improved. 

The accuracy of semi-empirical methods varies during cloudy periods when models might not differentiate clouds' impact on irradiance. However, the methods typically require less computational demand than physical models, so they can be applied to a wide region where calibration might be needed.

\subsection{Site-adaptation techniques}
\label{sec:siteadapt}

The site-adaption technique adjusts satellite-derived GHI by incorporating available ground measurements to provide estimates at a target location. The correction techniques involve linear multiple regression, distribution mapping (quantile or kernel density distribution), adjusting atmospheric parameters used in the satellite retrieval process, and machine learning techniques; more details in \cite{Polo2016, Polo2020}. 

\cite{Han2022} acquired Himawari-8 satellite images from JAXA and historical data of MERRA-2 product as inputs for REST-2 clear-sky model. The GHI estimation improvement relied on adjusting polynomial coefficients in the Heliosat algorithm, cloud albedo selection and linear bias removal as the site adaptation technique. The results were tested with hourly-averaged ground GHI in Taiwan with MAE of 38.74 $\wm$ and RMSE of 57.42 $\wm$.

Site adaptation techniques using nonlinear models from machine learning approaches can be found in \cite{Salamalikis2022}. The estimated GHI from the Copernicus atmospheric monitoring service (CAM-Rad) service and then corrected the value by calculating a perturbation from ML methods using solar zenith angle, AOD, water vapor, ozone, and cloud information as inputs. Overall, the two best performances were obtained by tree-based models (XG boost and random forest), which yielded an RMSE of 56-57 $\wm$. \cite{Verbois2023} used estimated GHI from HelioClim3, clear-sky index, and zenith angle as inputs for gradient boosting model to correct in both spatial and temporal domain.

Neural network techniques, presented in the next section, can capture a complex relationship between the satellite data and the corresponding ground-based GHI measurements.

\subsection{Neural network techniques}
\label{sec:nn}

Deep learning (DL) has become a leading trend in solar energy estimation due to its data-driven nature. It enables the parameters in an artificial neural network (ANN) to be fine-tuned according to the collected data; resulting in less manual customization of input features compared to the previous approaches that rely on the physical and meteorological information. 

Satellite data is the key to extending the coverage of solar forecasting to entire countries. However, the database stations used as the reference for implementing and evaluating solar forecasting are often relatively sparse (\eg,~\cite{Alobaidi2014} use only five stations), and the operation and installation can be costly.~\cite{Srivastava2018} address this limitation by using satellite-based GHI measurements in training, which offer the global irradiance information in Europe and USA; and this work shows that LSTM has the best performance among ANNs but is at a similar level to gradient boosting regression (GBR). Similarly, \cite{Yu2019} tested LSTM on the NSRDB database (USA). Nevertheless,~\cite{Qin2018} found that an ANN can still be outperformed by a physics-based model with high-quality inputs, \eg, Yang hybrid model with meteorological and MODIS satellite data, while cloud fraction and solar zenith angle were found the parameters that highly includes the model accuracy. 
 
Subsequent works further utilize the spatial information from satellite data as the reference to improve the model learning, \eg,  solar radiation images in~\cite{Yeom2020} and satellite-derived GHI in~\cite{Nielsen2021}. The ground measurements are discontinuous and sparse, which can be inadequate to represent the regional characteristic of the estimated GHI; see~\cite{Jiang2019}. Utilizing the spatial information, \eg, cloud mask~\cite{Nielsen2021}, spectral images~\cite{Jiang2019, Yeom2020}, infrared images~\cite{Ajith2021}, and satellite images~\cite{gallo2022solar}, as inputs can promote the continuity. To process the spatial information, these works~\cite{Jiang2019, Yeom2020, Ajith2021} attach convolution neural networks (CNN) for image feature encoding. Advanced CNN architectures are used as the IrradianceNet~\cite{Nielsen2021} and as the STUNet ~\cite{gallo2022solar} that employed ConvLSTMs and 3D-CNNs as image encoders, respectively. The time-series dynamics between satellite images are used in addition to the spatial information as an additional set of inputs. In such cases, LSTM-based models are often used to derive the forecast irradiance in the future timestamp~\cite{Ajith2021, Nielsen2021}.     
Fusing sky and satellite images can improve forecasting accuracy~\cite{ PalettaOmnivision2023}, which outperforms other methods that use only sky or satellite images. However, the sky-image based methods ~\cite{FengSolarNet2020, PalettaDL2021, PalettaECLIPSE2022, Liu2023, Zhang2023, PalettaOmnivision2023} focus on single-site forecasting only, which is not suitable for providing the GHI estimates at national level.

Although these satellite-based methods~\cite{Qin2018, Srivastava2018, Jiang2019, Yeom2020, Ajith2021, Nielsen2021, gallo2022solar} are promising, they are not applicable in our settings. Specifically, our system promises near real-time solar forecasting for 93,061 grids across Thailand, in which the system first acquires the satellite images from Himawari-8 provided at 10 minutes resolution, and then the satellite-derived GHI estimates are calculated and used to generate the solar map for every 30 minutes. As such, the target data-driven model has to offer a good trade-off between accuracy and efficiency. Thus, our work departs from the previous methods in the following aspects: 
\begin{itemize}
    \item \textit{Input data and parameters.} Unlike previous studies~\cite{Qin2018, Jiang2019} that rely on a number of terrain and atmospheric parameters, our model relies primarily on cloud data, with time-based features and the altitude parameter.    
    
    \item \textit{Learning reference.} Satellite-derived GHI from a physical-based method is used as the reference in model learning such as~\cite{Srivastava2018, Yeom2020, Nielsen2021}, but these references are rather indirect compared to the ground GHI measurements. Our setting tunes the model towards the GHI ground measurements.
    
    \item \textit{Runtime requirement.} Unlike the CNN-based methods~\cite{Jiang2019, Yeom2020, Ajith2021, Nielsen2021, gallo2022solar} whose CNN operation performs window sliding in each grid, we try to avoid such operation, not only because of the runtime requirement, but also to retain the details of each grid area that is roughly the size of a sub-district in Thailand  ($\approx4\sqkm$).   
\end{itemize}

Our key contributions are as follows.
\begin{enumerate}
\setlength{\itemsep}{5pt}
\item A clear-sky model for Thailand's weather condition. We improve the Ineichen \& Perez clear-sky mode  by re-estimating the Linke Turbidity using nonlinear regression. This is achieved by fitting clear-sky irradiance curve where $T_L$ is parameter using the collected ground measurements of GHI from 53-station data. 

\item  A methodology that exploits deep-learning or tree-based models to capture a complex relationship between satellite cloud and GHI. To find the best model, we benchmark LSTM, Informer, Transformer, and LightGBM (LGBM) for GHI estimation. Additionally, we improved Informer and Transformer particularly for this application. For LSTM and LGBM, we study the impact of input features. Our findings indicate that all these models exhibit overall performance comparable to commercial service X, a widely recognized industry benchmark. Our implementation are provided at~\url{https://github.com/energyCUEE/solarmap-DL-2024}.

\item A framework that operates in almost real-time implementation (30 minutes) and displays the estimated results across Thailand. Additionally, we provide the real-time solar energy map from the estimated irradiance map at hourly, monthly, and yearly resolution. This solar map can be useful for planning and operating solar energy production.    
\end{enumerate} 

\begin{figure}[ht]
\centering
\begin{subfigure}[b]{0.43\linewidth}
\centering
\includegraphics[height = 4cm]{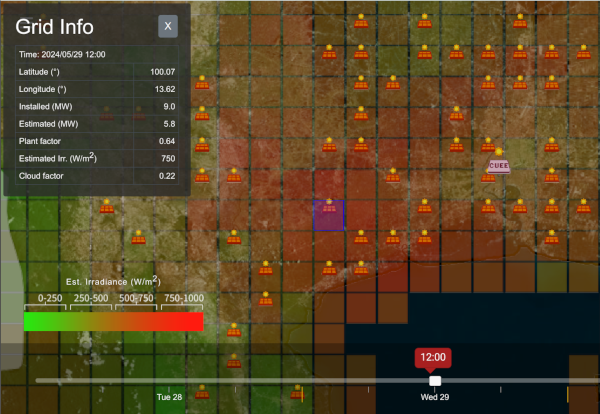}
\caption{Grid information}
\end{subfigure} \hfill
\begin{subfigure}[b]{0.55\linewidth}
\centering
\includegraphics[height = 4cm]{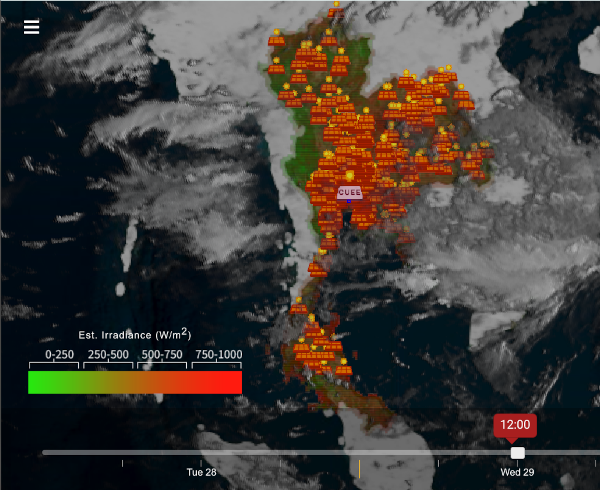}
\caption{Registered solar sites} 
\end{subfigure} \\
\begin{subfigure}[b]{0.32\linewidth}
\centering
\includegraphics[width=\linewidth]{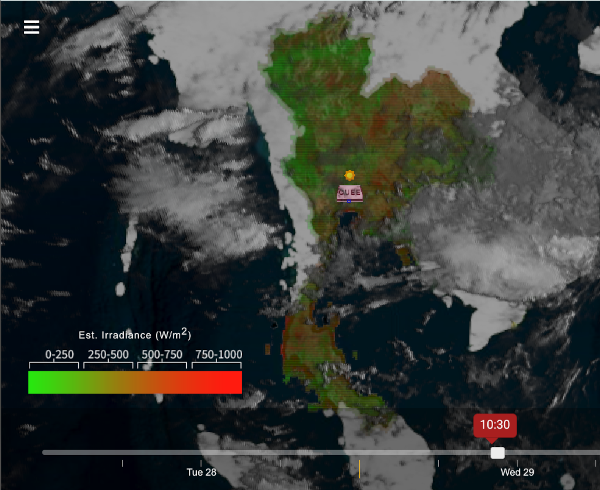}
\caption{at 10:30} 
\end{subfigure}
\begin{subfigure}[b]{0.32\linewidth}
\centering
\includegraphics[width=\linewidth]{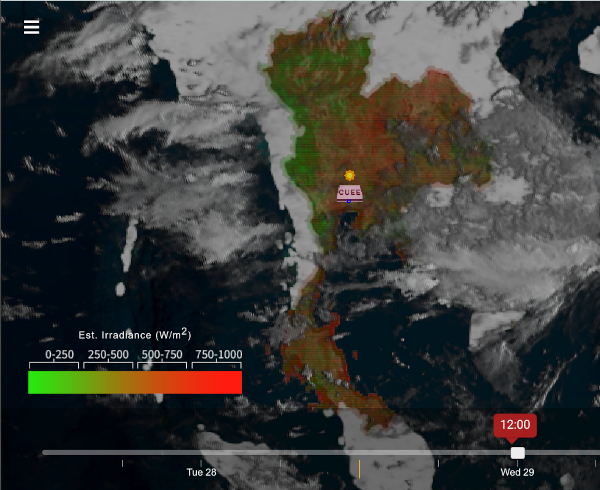}
\caption{at 12:30} 
\end{subfigure}
\begin{subfigure}[b]{0.32\linewidth}
\centering
\includegraphics[width=\linewidth]{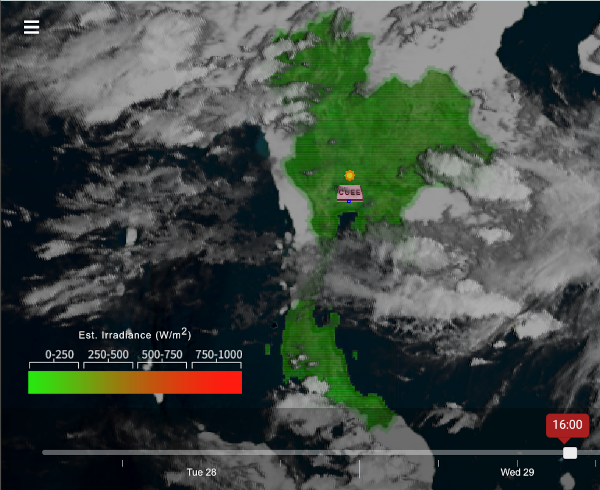}
\caption{at 16:00} 
\end{subfigure}
\caption{Thailand solar map displayed on \url{www.cusolarforecast.com}: (a) Estimated irradiancde and generated power on a selected grid are shown. (b) There are 576 registered solar sites with known installed capacity, as shown in the orange icons. (c)-(e) Solar map displays the estimated ground irradiance at various times.}
\label{fig:cuee_solarmap}
\end{figure}

\section{Thailand solar map}
Our Thailand solar map platform is currently running at \url{https://cusolarforecast.com/}, where the methodology is a machine learning model that takes input of cloud index, geographical location, and clear-sky irradiance to produce the GHI estimate. The random forest model was implemented from 2022 to June 2024. Our continuing plan of updating the model has been discussed in \cite{solarmappmap2024}, and the goal of this paper is to improve the model, considering other tree-based and deep-learning models. The service is currently running under the deep learning model, the Informer, as to be presented subsequently in \Cref{sec:ghiestim}. This section describes the map requirement and the available input data resources for developing the GHI estimation model. 

\subsection{Specification}
Thailand solar map is required to provide the following features to support the service displayed in \Cref{fig:cuee_solarmap}.
\begin{enumerate}
\item The map displays at $2.4 \times 2.4 \;\sqkm$ resolution within Thailand boundary, and contains 93,061 grids. The map intensity shows the estimated GHI that accounted for cloud information across Thailand and updates every 30 minutes.
\item The PV icons mark the locations of installed solar power plants, while the CUEE icon is our solar rooftop $8\;\kw$ at our research institution (Faculty of Engineering, Chulalongkorn University). The system shows estimated solar power and the cloud index amount extracted from Himawari-8 satellite images at the grids with solar sites installed.
\end{enumerate}

In this solar map, we use the information on solar power installed capacities and locations, published on the website of the Office of Energy Regulatory Commission (ERC) at \url{https://www.erc.or.th/en}, together with the estimated GHI from our models being updated every 30 minutes by using the cloud images received from Himawari-8 satellite, to estimate solar generated power across Thailand. The current updating frequency of 30-minute intervals is chosen to be consistent with the time resolution of the Unit Commitment program run by TSO. The solar map will also be designed to give convenience to aggregate estimated solar energy within a specified region, for example, in each control region, sub-region of TSO, or substation service area of DSO, employing spatial ensemble techniques thereafter.

\subsection{Data resources}
\paragraph{Ground GHI solar sites} There are 576 solar sites registered to the Energy Regulatory Commission of Thailand (ERC), where information on locations and installed capacity are provided. Ground measurements of GHI and satellite cloud data are required for model training and validation among these sites. \Cref{fig:solarsite} (b), shows 53 solar farms that record 15-minute average of solar irradiance, solar power, ambient and PV module temperature measurements from October 2020 to June 2023. The pyranometer model of solar irradiance measurement is SMP 6 (ISO 9060 spectrally flat class B). To ensure data quality, a pre-processing pipeline was implemented. 

Initially, data was filtered to retain only daytime measurements. GHI values were restricted to a physically realistic range of 0 to 1400 $\wm$. Dates with excessively low average daytime GHI were discarded. This step aimed to eliminate periods with insufficient solar radiation or inaccurate measured values. Dates with missing data periods exceeding six hours were entirely removed. Shorter data gaps were filled using linear interpolation. Last, a manual inspection of GHI data and associated variables for consistency was performed. As a result of this quality control process, three of the original 56 sites were unsuitable and excluded from the analysis. The remaining 53 sites were deemed sufficiently reliable for this study.

\begin{figure}[ht]
\centering
\includegraphics[width=0.8\linewidth]{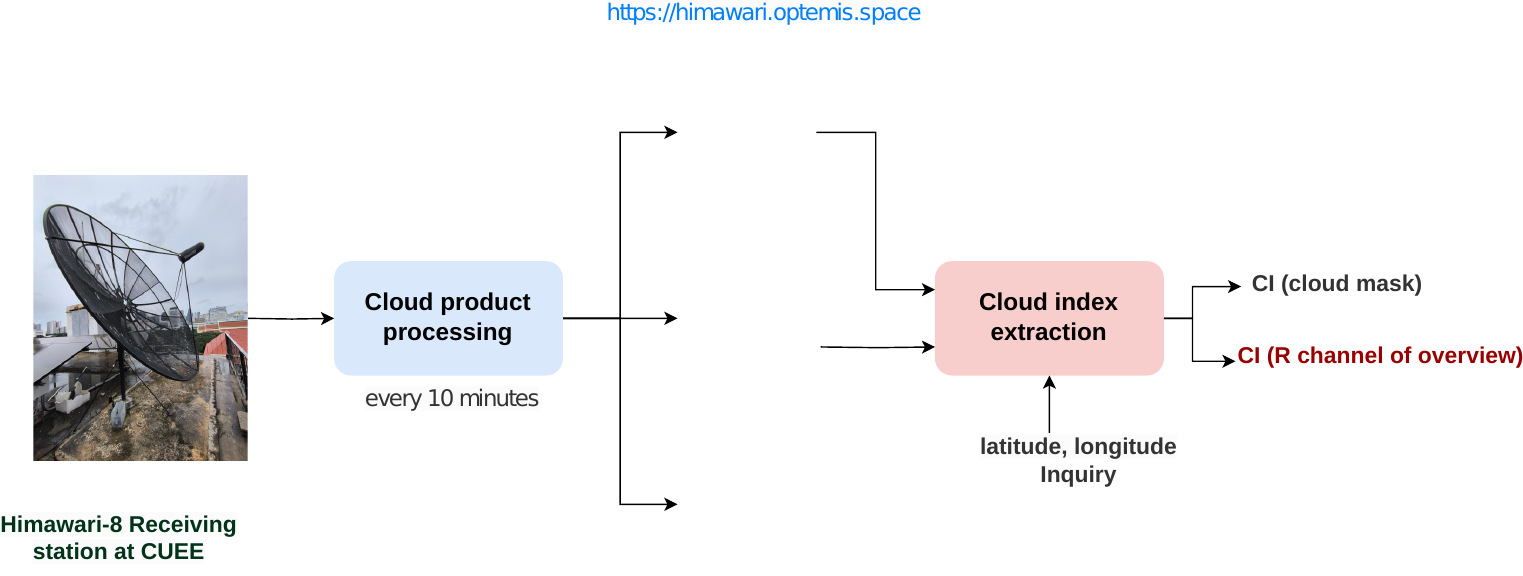}
\caption{Himawari-8 receiving station at CUEE and cloud index processing system.}
\label{fig:satellite_dishCUEE}
\end{figure}

\paragraph{Himawari-8}  
The 8th Himawari geostationary weather satellite operated by the Japan Meteorological Agency (JMA) provides 16-channel multispectral images that capture visible light and infrared images of the Asia-Pacific region. A C-band satellite dish and receiving station have been installed at our research institution, CUEE, since February 2022; see \Cref{fig:satellite_dishCUEE}. There are three types of cloud products: composite images, cloud masks, and colorized images available at \url{https://himawari.optemis.space/archives}. We employ cloud mask images and the R-channel of the composite image that has been processed for sun zenith correction and enhanced for smoothness. The R, G, and B channels correspond to 0.6, 0.8, and 10.8~$\mathrm{\mu m}$ wavelength, respectively. Both the cloud mask and R-channel images have $1670\times1725$ pixels with the resolution of $2\times2~\sqkm$ covering  Thailand area (\approxtext~20.6-5.4\degreeNorth, 97.1-106\degreeEast), and are released every 10 minutes. These images in TIFF format contain geographical information. Given an inquired latitude and longitude, we can extract the cloud intensity at the corresponding pixel and convert it to cloud index as $\ci_M$ (from cloud mask) and $\ci_R$ (from R-channel) as follows. 

The data during the period of satellite outage or error image scan are flagged to be imputed. If the imputation period is small, a simple interpolation from nearby data points is performed, while the missing data in a long period are removed from the study. 

\begin{figure}[h!]
\includegraphics[width=0.95\linewidth]{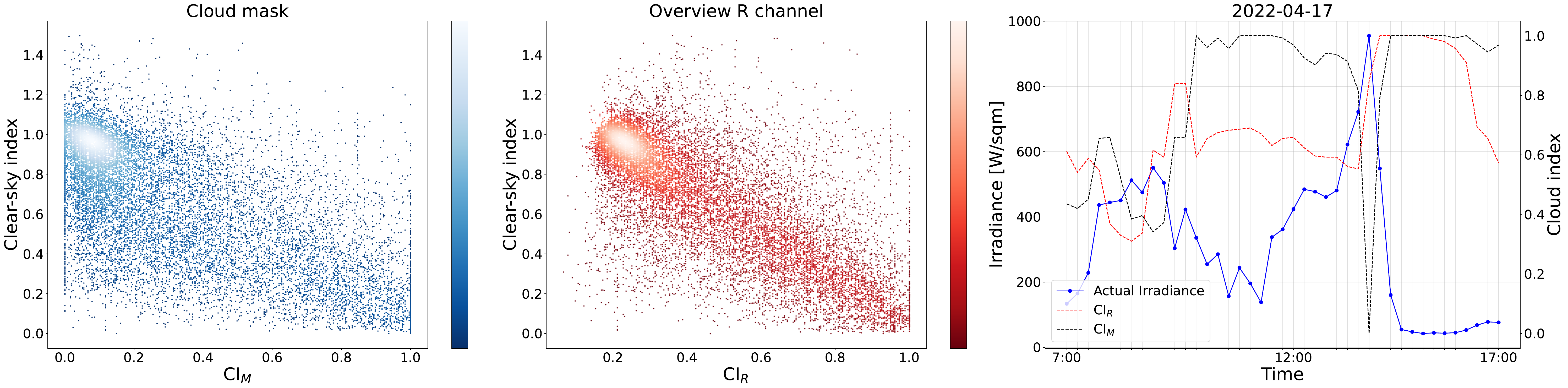}
\caption{Cloud index computed by \cref{eq:cloud_index}. The GHI time series drops to small values when the cloud index is high.}
\label{fig:scatter_CI_k}
\end{figure}

\paragraph{Cloud index extraction} In \cite{Paulescu2013}, the cloud index is calculated by 
\begin{equation}
\ci = \frac{X - \mbox{LB}}{\mbox{UB} - \mbox{LB}},
\label{eq:cloud_index}
\end{equation}
where $X$ is the pixel intensity of interest (0-255); UB and LB are the upper and lower bound of the dynamic range at a given point in time and space. In our experiment, LB is zero, and UB is 255. The cloud indices computed from cloud mask, $\ci_M$, and R channel, $\ci_R$, were plotted versus the clear-sky index $k(t) = I(t)/\iclr(t)$ shown in \Cref{fig:scatter_CI_k}, suggesting that $\ci_R$ has a stronger anti-correlation with $k(t)$ than $\ci_M$. 

Pre-processing steps were applied to cloud images and extracted cloud index (CI) data to ensure alignment with the target variable, GHI measurements. Initially, satellite images containing significant areas of bad scan (black area) were discarded. For each day, the first and last cloud image within the 7:00 to 17:00 timeframe were identified. Missing cloud images within this range were imputed using a forward-fill method. To accommodate the GHI estimation model's requirement for five lagged CI values, a backward-fill approach was employed for the first five time steps. For the imputed date-time between 5:00 and 6:00, zero-filling was applied; otherwise, backward-filling from the first image's CI value was utilized.

To train and infer a GHI estimation model, we must extract $\ci$ at a pixel of cloud images that best represents the desired location using GeoTIFF data to find the nearest pixel whose coordinate (latitude /longitude) is closest to the coordinate of interest. 

\paragraph{Re-analysis weather data from NWP (numerical weather prediction) model} We currently subscribe to the \emph{Meteo Monitoring} and the \emph{Meteo historical} services from \url{https://www.soda-pro.com/}. The Meteo monitoring service provides day-ahead GFS weather forecasts by the National Centers of Environmental Prediction (NCEP). At the same time, the Meteo historical contains reanalysis data of solar irradiance and temperature estimates (denoted as $I_\mathrm{nwp}$ and $T_\mathrm{nwp}$) with a resolution of 15 minutes retrieved from the MERRA-2 model (Modern-Era Retrospective analysis for Research and Applications). For training, we incorporate the MERRA-2 GHI and temperature as part of the features for GHI estimation models at 53 solar locations.

 Cloud data, NWP data, and GHI measurements overlapped from April 2022 to June 2023, resulting in our dataset of 22,906 days or 1,195,257 samples. 

\section{Methodology}
\subsection{Ineichen clear-sky model}
\label{sec:clearsky_model}
A clear-sky model provides a GHI value at a given time and location under a clear-sky assumption. It typically takes date-time to estimate the solar zenith angle or the sun position and uses the location's coordinate to acquire the altitude and other atmospheric parameters. Benchmarking of 75 clear-sky models can be found from \cite{Gueymard2019, Yang2020, Sun2021,Bright2020}, where it was reported that the REST2 model is among the best-performing models in the equatorial zone. At the same time, it requires several atmospheric parameters such as aerosol optical depth, total ozone, nitrogen dioxide amounts, or aerosol single-scattering coefficient. Among these 75 models, simplified-Solis  model\cite{IneichenSolis2008}, Kasten, and Ineicen \& Perez models require fewer numbers of physical input parameters. 

In this paper, we consider the Ineichen \& Perez clear-sky model \cite{Ineichen2002} due to its simplicity in acquiring only one atmospheric parameter, the Linke turbidity. The model is described by
\begin{equation}
\iclr(t) = a_1 I_0 \cos \theta(t) e^{-a_2 \am(t)( f_{h_1}+ f_{h_2} (T_L -1) )}.
\end{equation}
The extraterrestrial irradiance constant, $I_0= 1366.1 \wm$ is the theoretical value of irradiance on the Earth's surface if the Earth were not surrounded by an atmosphere. The altitude of a desired location, $h$,  is used for computing the constants:
\[
f_{h_1} = e^{-h/8000},\; f_{h_2} = e^{-h/1250},\; a_1 = 5.09 \times 10^{-5}h + 0.868,\; a_2 =  3.92 \times 10^{-5} h +0.0387,
\]
where $h$ is retrieved from STRM30 dataset \cite{srtm}.
We follow the model of airmass coefficient, $\am(t)$, from \cite{Paulescu2013} given by
\begin{equation}
\am(t) = \frac{1}{ \cos \theta(t) + 0.50572( 96.07995 - \theta(t))^{-1.6364}}.
\end{equation}
The Linke turbidity parameter $T_L$ represents the amount of atmospheric absorption by water vapor and aerosol particles that attenuate extraterrestrial radiation. This factor can be obtained via experiments on clear-sky days. Still, it is more common to be empirically estimated \cite{Eltbaakh2012}, where $T_L$ is empirically estimated to best match with local ground measurement of clear-sky irradiance in Thailand. 

\begin{figure}[ht]
\centering
\begin{subfigure}{0.44\linewidth}
\centering
\includegraphics[width=1\linewidth]{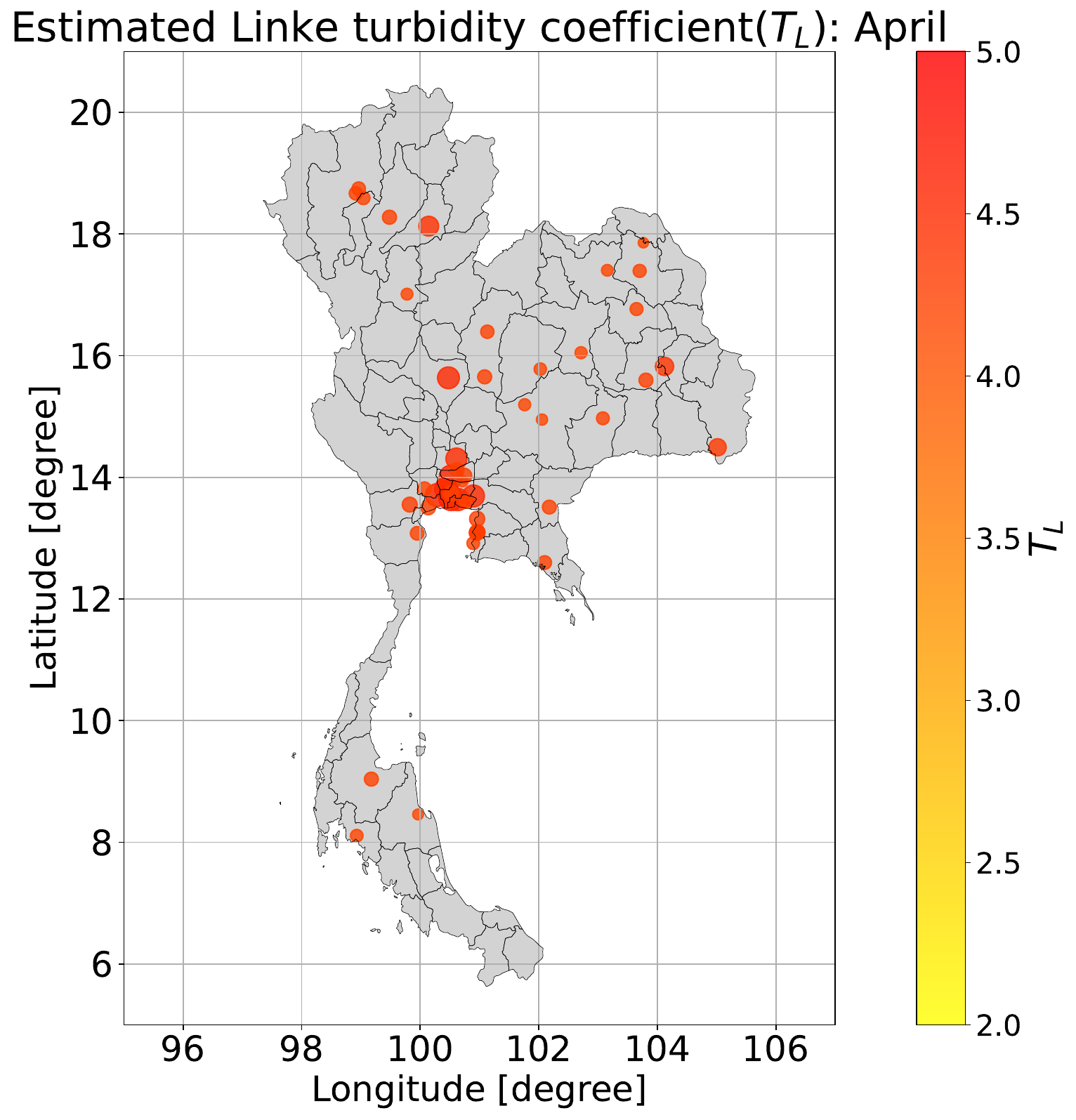}
\caption{Estimated coefficients $T_L$ of April across 53 locations.}
\end{subfigure}
\begin{subfigure}{0.47\linewidth}
\centering
\includegraphics[width=1\linewidth]{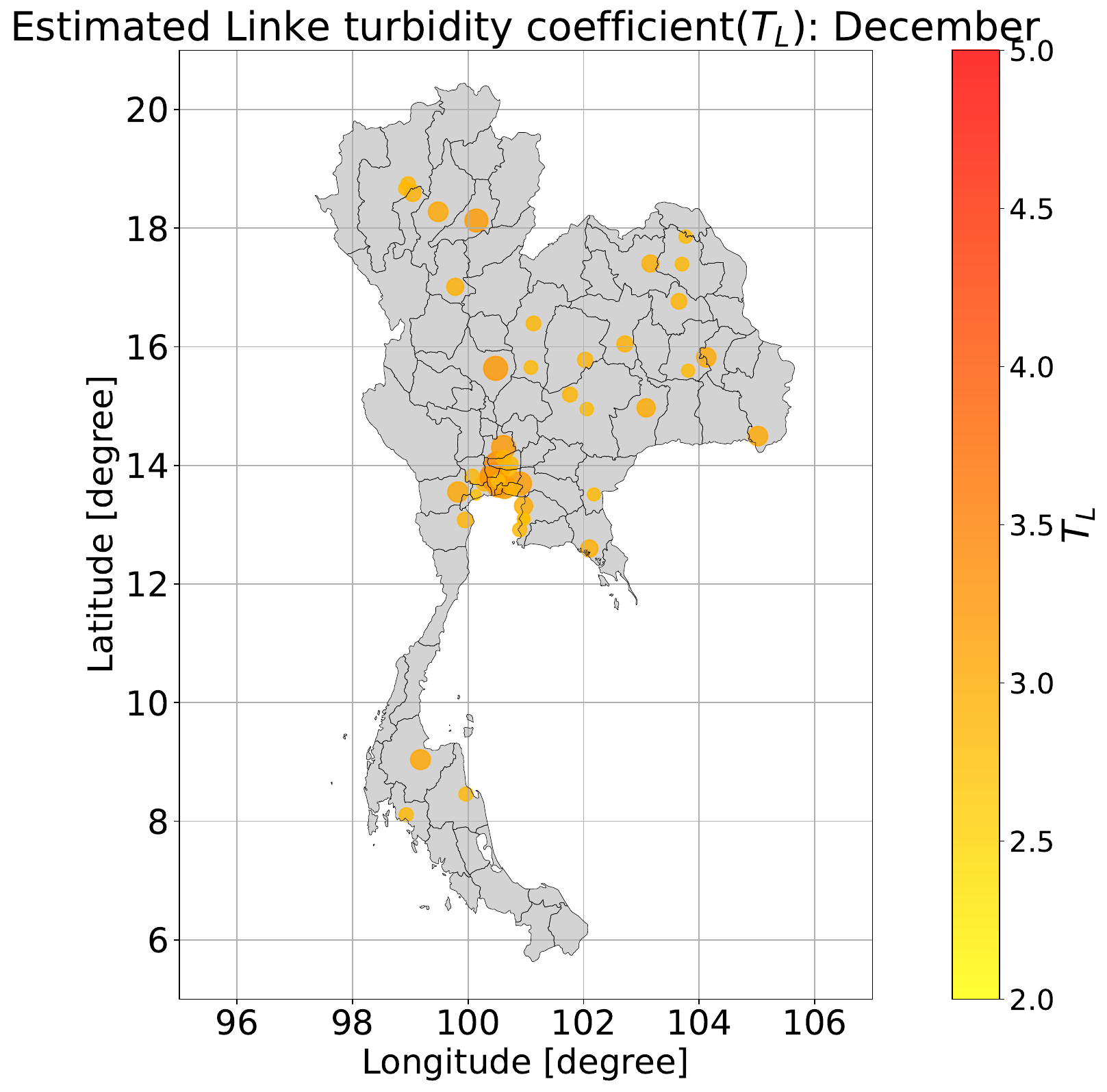}
\caption{Estimated coefficients $T_L$ of December across 53 locations.}
\label{fig:TLmap}
\end{subfigure} \\
\begin{subfigure}{0.85\linewidth}
\centering
\includegraphics[width=0.8\linewidth]{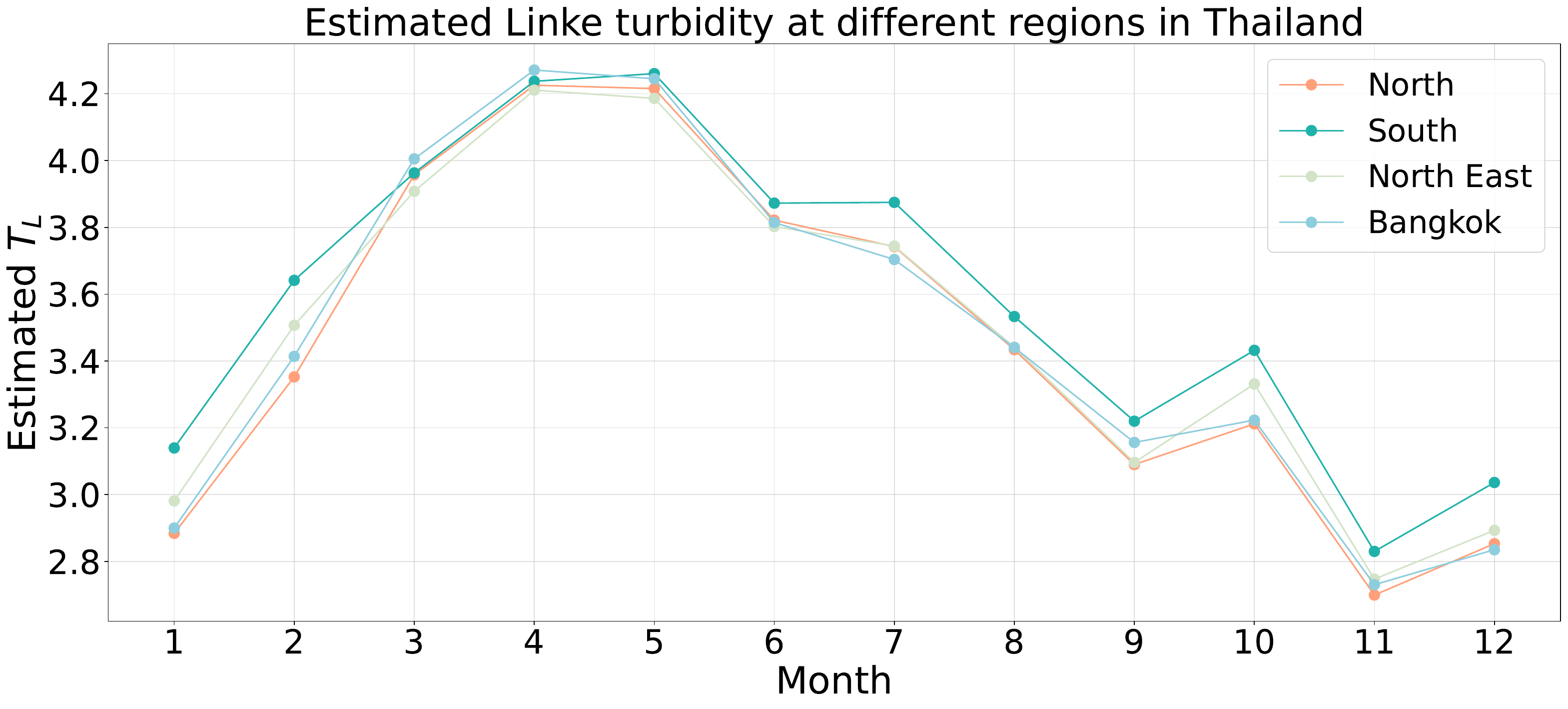}
\caption{Estimated coefficients $T_L$ across 12 months.}
\label{fig:TLmonth}
\end{subfigure}
\caption{Results of Linke turbidity coefficients ($T_L$) estimation of 12-month data.}
\label{fig:TL}
\end{figure}

We have set up an experiment to adjust the Linke turbidity coefficients based on ground measurements of GHI from 53-station data. The clear-sky GHI portions are extracted from PVlib clear-sky detection implementation~\cite{pvlib2023} and split into train:test at ratio 90:10. The training set is divided into 12 months and monthly $T_L$ coefficients are estimated separately since the value $T_L$ is known to vary seasonally~\cite{Ineichen2008, Garniwa2023}. The estimation technique is nonlinear regression to fit the clear-sky GHI curves. \Cref{fig:TLmap} shows the result of estimated $T_L$ as a heatmap displaying the value across the Thailand map, where its value is high during April but appears lower in December. A complete trend of $T_L$ estimated monthly is shown in \Cref{fig:TLmonth}. The coefficient $T_L$ is at peak during Apr-May and smaller in other months.

Other guidelines of $T_L$ estimation in Thailand can be found in \cite{Janjai2003,Chaiwiwatworakul2004}. Other papers about the Linke estimation are presented in \cite{Ineichen2008, Garniwa2023}.

\subsection{Satellite-derived GHI estimation}
\begin{figure}[h!]
\centering
\includegraphics[width=0.8\linewidth]{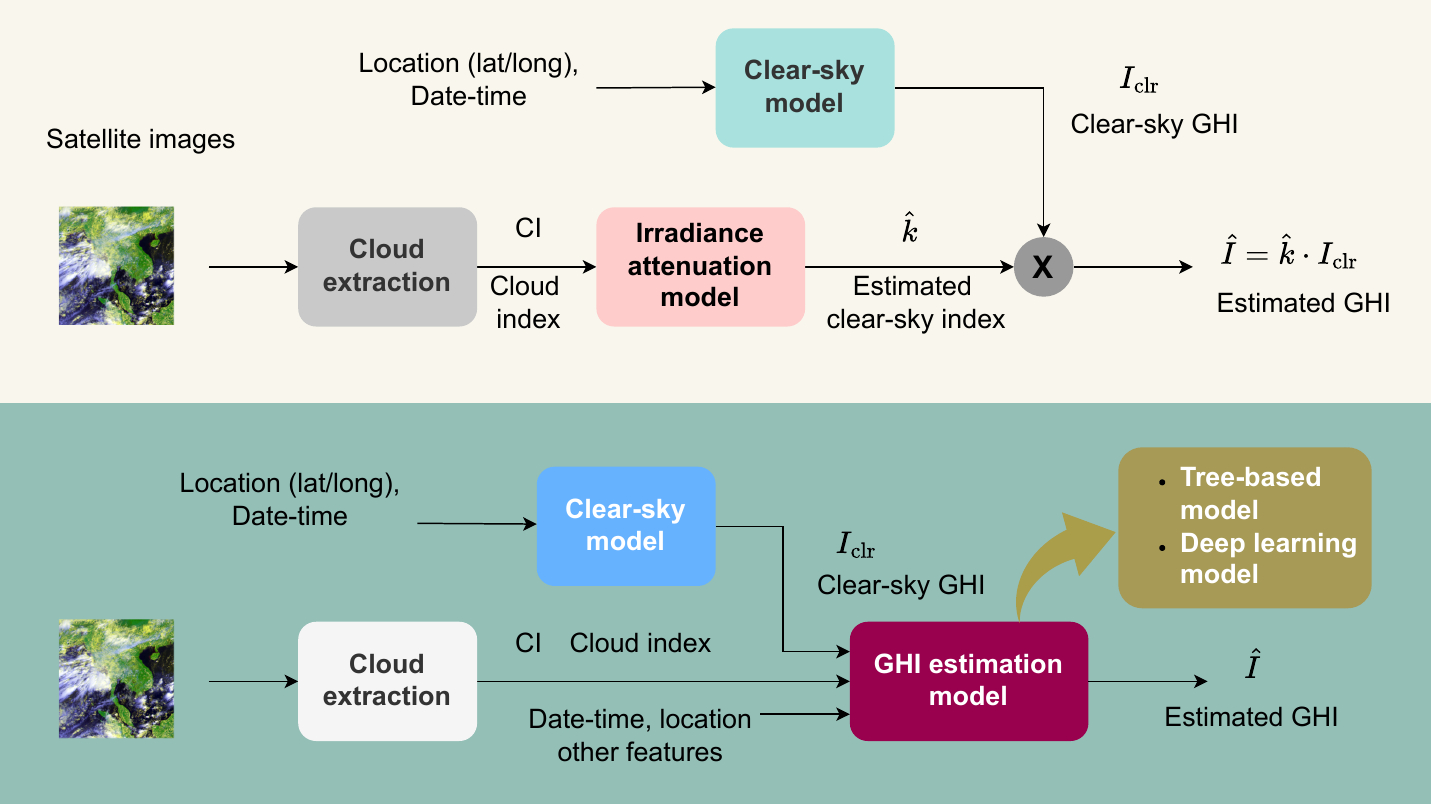}
\caption{GHI estimation schemes: \emph{(Top.)} Physic-based or empirical method. \emph{(Bottom.)} Proposed scheme.}
\label{fig:model_diagram}
\end{figure}

The estimation of GHI based on satellite-derived information has two schemes explained in \Cref{sec:intro}. The physics-based and semi-empirical methods are typically based on the top diagram in \Cref{fig:model_diagram}. The cloud index extracted from satellite images is an input for an irradiance attenuation model that produces an estimate of clear-sky index. Examples in literature are Heliosat algorithms \cite{Rigollier2004,Qu2017}, piece-wise polynomial and rational functions~\cite{solarmappmap2024}. The estimated $\hat{k}$ is then used for compensating a reduction in irradiance through the relationship $\hat{I} = \hat{k} \cdot \iclr$.  

Instead of expressing the effect of the cloud on the clear-sky index directly, we propose to use a nonlinear mapping as \textbf{GHI estimation model} to associate cloud index, clear-sky GHI, and other features with the estimate GHI, as shown in the bottom diagram of \Cref{fig:model_diagram}. The non-linear map can better capture complex cloud effects on solar irradiance, and various choices of GHI estimation model can be considered, \eg, tree-based models were validated with a good MAE/RMSE performance in \cite{solarmappmap2024}.  Note that the input features for the non-linear mapping are all meteorological variables, especially the cloud index, but not GHI measurements as they are not available spatially across Thailand when deploying the model in practice. The list of features we used in this paper are
\[
\ci_M, \ci_R, \iclr, I_{\mathrm{nwp}}, T_{\mathrm{nwp}}, \lat, \llong, \hour, \dday
\]
\noindent Let $x(t) := \Set{\ci_M, \ci_R, \iclr, I_{\mathrm{nwp}}, T_{\mathrm{nwp}}, \lat, \llong, \hour, \dday}$ denotes a vector collecting these input features at time $t$. Then, the input sequence $\FeatureSeq$ is  the collection of the input $x(t)$ and its lags, \ie,   
\[
\FeatureSeq = \RBrc{x(t), x(t-1),\ldots, x(t-\seqlen)}  
\]
\noindent where $\seqlen$ is the length of input sequence. Suppose that $\Model_{\Theta}(\cdot)$ is a GHI estimation model that is data-driven with tuned parameters $\Theta$. The GHI estimate at time $t$ can be obtained by performing the model inference: $\Predicted (t) = \Model(\FeatureSeq; \Theta)$. 

When training a model, it is natural to enforce the similarity between the prediction $\Predicted (t)$ and the ground truth $\GroundTruth (t)$. In this study, we benchmark the following methods: \textbf{LightGBM (LGBM), LSTM, Informer, Transformer, and the service X}. For service X's predictions, we have downloaded historical GHI estimates at 53 stations during the period of this study. We modified Transformer and Informer to acquire time-series data for solar irradiance forecasting while building upon existing LSTMs with additional linear layers. The hyper-parameters of these models were tuned and set with the values shown in \Cref{tab:param}.

\begin{table}
\tablefontsize
\centering
\caption{Benchmarked GHI estimation models.}
\begin{tabular}{ll} \hline
\textbf{Model} & Description \\ \hline
LGBM & \texttt{max\_depth} = 15, \texttt{p}=5, \texttt{num\_leaves} = 31, \texttt{num\_iter} = 500, \texttt{learning\_rate} = 0.1  \\ 
LSTM & \texttt{L}=1, \texttt{p}=5, \texttt{hidden}=128, \texttt{Dropout}=0.001, \texttt{learning\_rate}  = 0.001 \\
Informer & \texttt{L}=2, \texttt{p}=5, \texttt{hidden}=16, \texttt{Dropout}=0.001, \texttt{learning\_rate}  = 0.001 \\
Transformer & \texttt{L}=2, \texttt{p}=5, \texttt{hidden}=64, \texttt{Dropout}=0.001, \texttt{learning\_rate}  = 0.001 \\  \hline
\end{tabular}
\label{tab:param}
\end{table}

\subsubsection{LightGBM}
\label{sec:LGBM}
 LightGBM~\cite{LightGBM_NIPS2017} (light gradient boosted machine) builds on the concept of gradient boosting decision trees. LightGBM uses  the leaf-wise approach that prioritizes growing trees in informative areas, leading to more accurate models with fewer trees. Also, it has a process to select important features and remove less impactful features from the data. In our case, we provide  the following features as the input to the LightGBM: 
\[
\ci_M(t-p),\ldots, \ci_M(t), \ci_R(t-p), ..., \ci_R (t), \iclr, I_{\mathrm{nwp}}, T_{\mathrm{nwp}}, \lat, \llong, \hour, \dday
\]
The parameter setting of LightGBM is provided in~\Cref{tab:param}.

\subsubsection{LSTM}
\begin{figure}[h!]  
\begin{minipage}[b]{.24\linewidth}
  \subfloat[LSTM network\label{fig:DNN:LSTM_overall}]{\includegraphics[width=0.8\hsize]{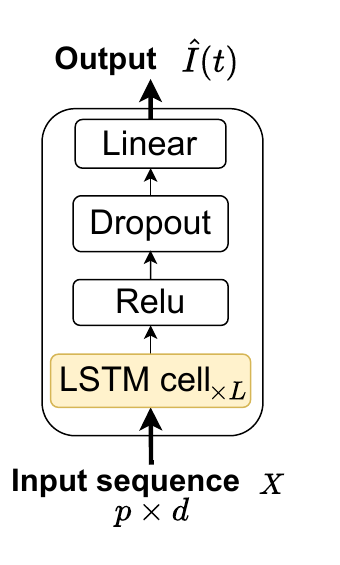}}  
 \end{minipage}
  \begin{minipage}[b]{.70\linewidth}
  \subfloat[LSTM cell~\label{fig:DNN:LSTM}]{\includegraphics[width=\hsize]{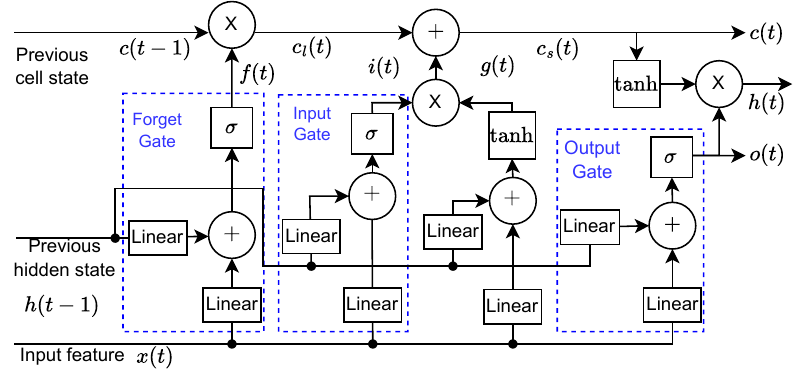}} 
  \end{minipage}
\caption{(a) LSTM network and (b) LSTM cell.}
\end{figure}

 Our LSTM (long short-term memory) neural network is enhanced from that of~\cite{Srivastava2018}  to include $L$ layers of LSTM cells, an activation, a dropout, and a linear layer, as shown in~\Cref{fig:DNN:LSTM_overall}, where  $\FeatureSeq \in \Real^{p\times d}$ is fed as the input, and the predicted irradiance $\Predicted(t) \in \Real$ is the output.  The LSTM cell is provided in~\Cref{fig:DNN:LSTM}. Its cell state and hidden state are initialized at zero. The LSTM cell recursively processes each input lag from $\FeatureSeq$. For an input $x(t-p)$, the long-term memory $c_l(t-p)$ is updated from the previous $c(t-p-1)$ and the forget gate $f(t-p)$. The short-term memory $c_s(t-p)$ is updated with the input gate $i(t-p)$ and $g(t-p)$. The output gate provides the output $o(t-p)$ and is used to update next hidden state $h(t-p)$. After each recursion, the output is collected as a latent sequence  $o(t-p),\ldots, o(t)$, fed to Relu activation, dropout layer, and linear layer. The linear layer maps the output sequence into a sequence of GHI estimates $\Predicted(t-p),\ldots, \Predicted(t)$, where the last element is the prediction $\Predicted(t)$.

\subsubsection{Transformer}
\begin{figure}[h!]  
\begin{minipage}[b]{.64\linewidth}
 \subfloat[Transformer\label{fig:DNN:Transformer}]{\includegraphics[trim={0 0.60cm 0 0.50cm},clip, width=\linewidth]{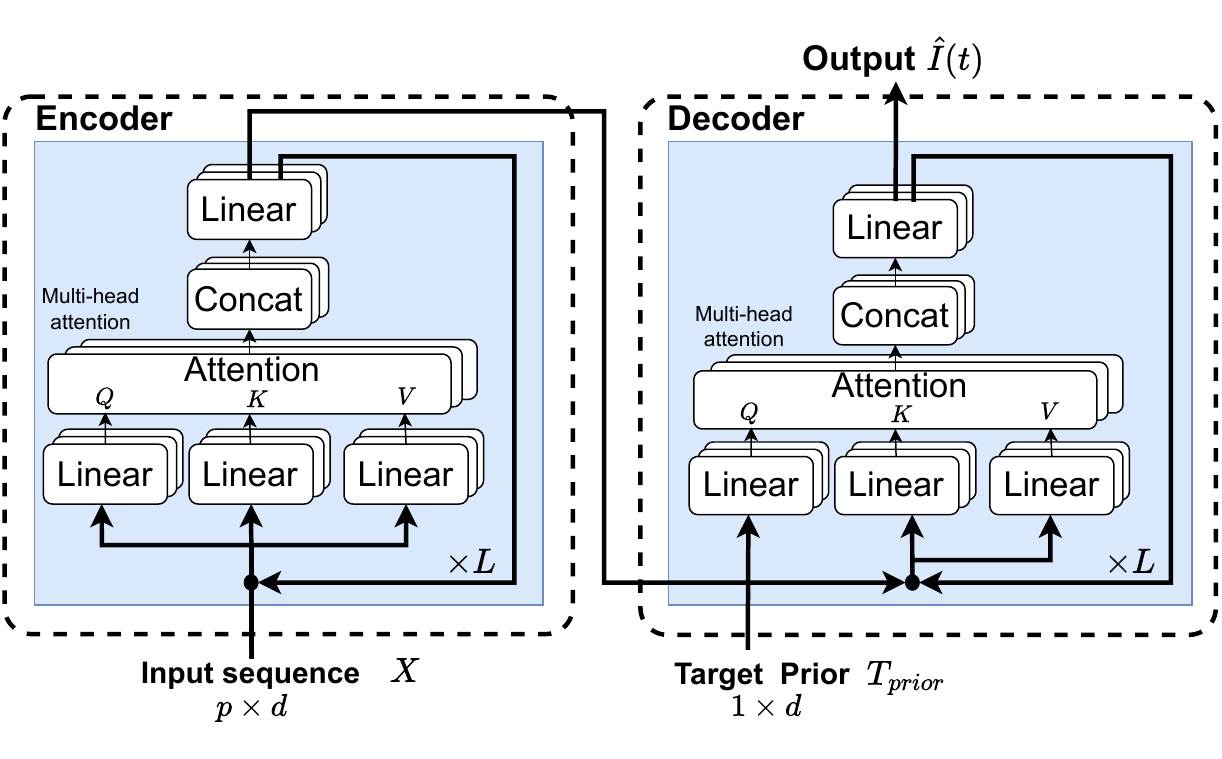}} 
 \end{minipage}
  \begin{minipage}[b]{.120\linewidth}
 \subfloat[Attention\label{fig:DNN:Attention}]{\includegraphics[width=\hsize]{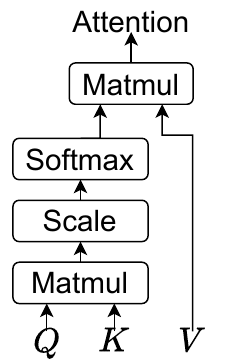}}
  \end{minipage}
   \begin{minipage}[b]{.170\linewidth}
  \subfloat[Sparse attention\label{fig:DNN:SparseAttention}]{\includegraphics[width=\hsize]{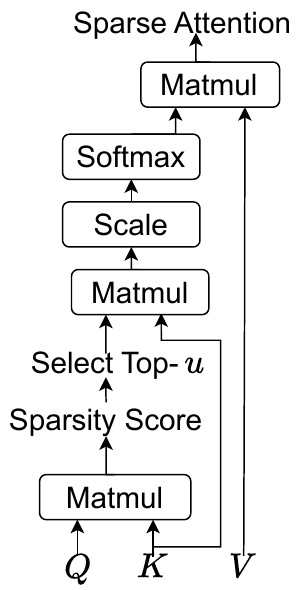}}
  \end{minipage}
\caption{(a) Transformer architecture,  (b) attention mechanism,  and (c) sparse attention mechanism.}
\end{figure}

  We modify the Transformer~\cite{Transformer_NIPS2017} by, \textit{first}, applying the attention mechanism to capture the dependencies within a time-series sequence and between the pairs of such sequences. \textit{Second},  the architecture is modified for the shorter sequence, resulting in the encoder-decoder pairs shown in~\Cref{fig:DNN:Transformer}, which is similar to~\cite{Informer_AAAI2020}, except that our input features and target priors are not the target itself.  At first, the encoder receives the input sequence $\FeatureSeq$, which calculates the self-attention to capture the dependencies within the sequence itself. Three linear layers (Linear) encode $\FeatureSeq$ as $Q$, $K$, and $V$ in $\Real^{p\times d}$. The self-attention mechanism can be expressed as follows (and in~\Cref{fig:DNN:Attention}):
\begin{align}
\selfAttention~(Q, K ,V)  =  \softmax\Brc{\frac{QK^T}{\sqrt{d}}} V \label{eq:DNN:Attention}     
\end{align} 
The decoder derives the prediction from cross-attention between target priors ($T_{prior}$) and the latent features from encoder. The target priors are the sequence of the following features:
\begin{align}
T_{prior} = \Set{\iclr, \lat, \llong, \hour, \mmonth, \dday,  I_{\mathrm{nwp}}, T_{\mathrm{nwp}}}\label{eq:DNN:TargetPrior} 
\end{align} 
\noindent which are used as the query in the attention calculation. The latent features from the encoder are the key-value pair.  

\subsubsection{Informer}
The architecture of Informer~\cite{Informer_AAAI2020} is similar to that of Transformer, except that the generic attention mechanism is replaced with the sparse attention mechanism~\cite{Liu2023, Informer_AAAI2020}.  We further modify Informer for predicting the solar irradiance $\Predicted(t)$ by adding the target priors as~\ref{eq:DNN:TargetPrior}.  At first, the score of the queries' sparsity is calculated, and then only the top-$u$ queries with the highest scores are selected to construct a sparse query matrix. Thus, we can construct a sparse query matrix (denoted as $\bar{Q}$) that contains only the top-$u$ highest sparsity scores. The sparse attention mechanism can be formulated by replacing  $Q$ with $\bar{Q}$ in~\Cref{eq:DNN:Attention}, \ie,   
\begin{align}
\sparseselfAttention~(Q,K,V) =  \softmax\Brc{\frac{\bar{Q}K^T}{\sqrt{d_k}}} V 
\label{eq:DNN:SparseAttention}     
\end{align}
\Cref{fig:DNN:SparseAttention} shows the sparse attention mechanism.

\section{Results and discussion}

\subsection{Data preparation}
Since cloud cover can vary greatly throughout the day,  traditional seasonal patterns become less reliable. It is important to ensure that the data samples in the training, validation, and test datasets share a similar degree of uncertainty. We classify GHI data of each day into three sky conditions: clear, partly cloudy, and cloudy, according to the rate of change in GHI and the daily average of clear-sky index $k(t)$, or $\bar{k}$. If the GHI curve has a smooth clear-sky curve, its slope must decrease (change from a positive value to zero and a negative value); meanwhile, the higher $\bar{k}$ indicates the lower cloud coverage. The criteria of our classification are described as follows: i) \emph{clear-sky:} GHI curve has decreasing slope, and $\bar{k} \geq 0.6$, ii) \emph{partly-cloudy:} GHI curve that are not clear-sky, and $\bar{k} \geq 0.6 $, and iii) \emph{cloudy:} GHI curve that are not clear-sky, and $\bar{k} < 0.6$.

The extracted cloud index and ground GHI data are merged into a synchronized 15-min resolution during 03:00-17:00 (local time in Thailand), totaling 1,195,257 samples or 20,924 days. The target forecasting hours range from 7:00 to 17:00. The data is divided with the ratio of 60:20:20 for train:validation:test, where each of the three sets is ensured to contain a uniform mix of three sky conditions see \Cref{tab:datasplit}.  To achieve this, we first segmented the data with lags and stacked them as sequences labeled with sky condition. Then, we shuffle the data and split them for training, validation, and testing. 

\begin{table}[h!]
\tablefontsize
\centering
\caption{Data split under sky conditions and data samples in train, validation, and test sets.}
\begin{tabular}{lrr|rrr} \hline
Sky condition & Total days  & Total samples &  Train & Validation & Test \\ \hline 
clear-sky & 921  & 49,038 & 28,979 & 10,907 & 9,152 \\ 
partly-cloudy & 21,549   & 1,123,391 & 675,227 & 224,739 & 223,425 \\
cloudy & 436  & 22,828 & 14,140 & 4,426 & 4,262 \\ \hline
Total & 22,906  & 1,195,257 & 718,346 & 240,072 & 236,839\\ \hline
\end{tabular}
\label{tab:datasplit}
\end{table}

\subsection{Clear-sky model evaluation}
\begin{figure}[ht]
\centering
\begin{subfigure}{0.90\linewidth}
\centering
\includegraphics[width=\linewidth]{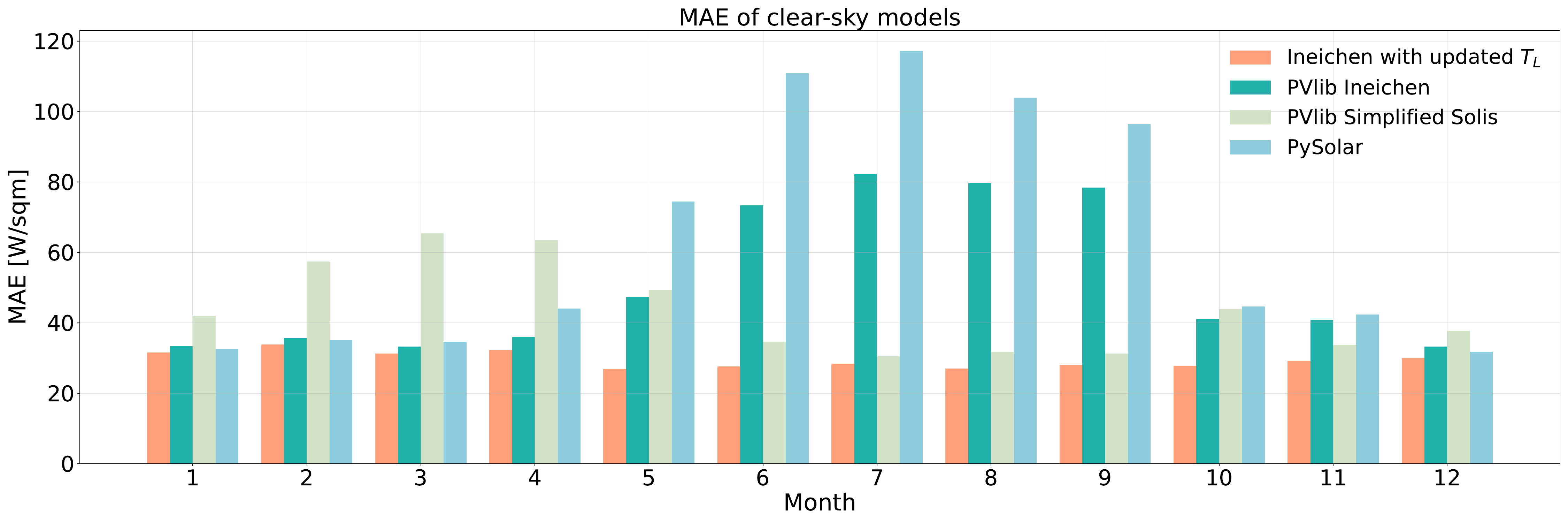}
\caption{MAE of clear-sky GHI.}
\end{subfigure}
\begin{subfigure}{0.90\linewidth}
\centering
\includegraphics[width=\linewidth]{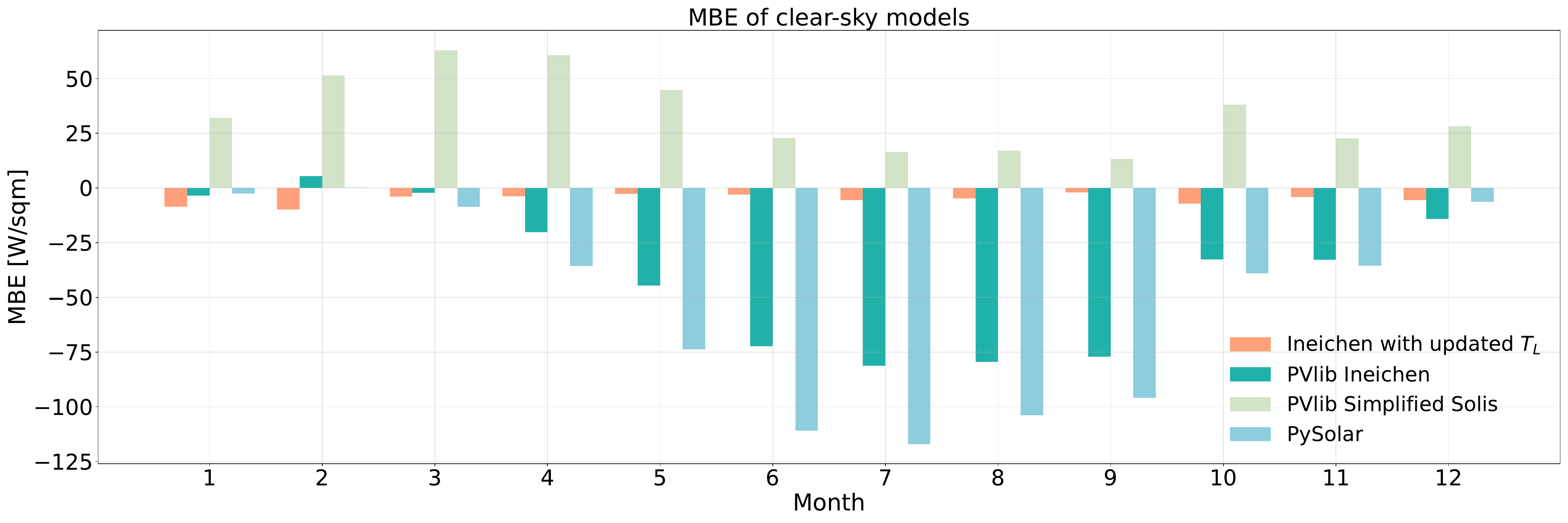}
\caption{MBE of clear-sky GHI.}
\end{subfigure}
\caption{Validated results of clear-sky models with clear-sky measurements.}
\label{fig:clearsky_bymonth}
\end{figure}

According to our improvement of the Ineichen clear-sky model in \Cref{sec:clearsky_model}, we benchmark the results against the following methods. 
\begin{enumerate}
\item Ineichen model with $T_L$ updated using clear-sky GHI measurements (proposed).
\item Ineichen model computed by PVlib~\cite{pvlib2023}.
\item Simplified Solis model computed by PVlib~\cite{pvlib2023}.
\item Pysolar \cite{pysolar}. The library does not provide GHI but only returns DNI, so we multiply DNI with $\cos \theta(t)$. 
\end{enumerate}
\begin{figure}[ht]
\centering
\includegraphics[width=0.85\linewidth]{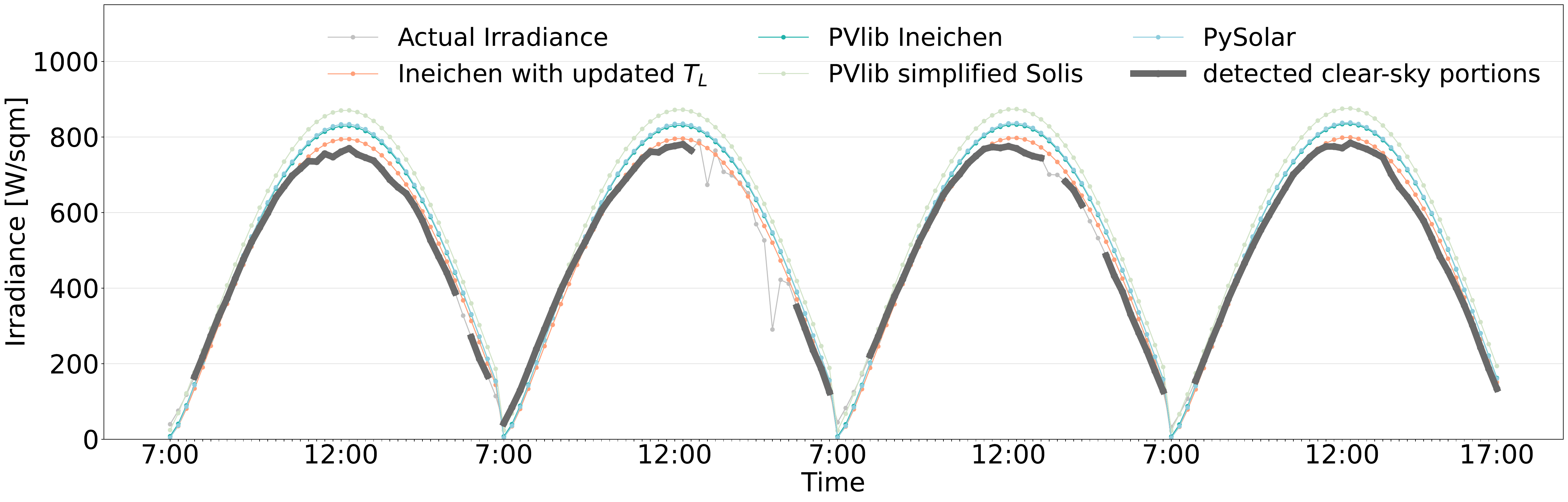}
\caption{GHI time series that are detected as clear-sky (thick gray portions.)}
\label{fig:clearsky_ts}
\end{figure}
\Cref{fig:clearsky_ts} shows that after we tuned $T_L$ to local GHI data in Thailand, the Ineichen clear-sky model fits better to clear-sky time series while other methods tend to be overestimated (on selected sample dates in \Cref{fig:clearsky_ts}. The MBE in \Cref{fig:clearsky_bymonth} (b) clearly shows that PVlib Ineichen and PySolar highly underestimate clear-sky GHI during Jun-Sep, while PVlib simplified Solis overestimates during Feb-May. Our method achieves the least MAE and MBE in all months.

\subsection{Satellite-derived GHI estimation}
\label{sec:ghiestim}

\begin{figure}[h!]
\centering
\includegraphics[width=0.9\linewidth]{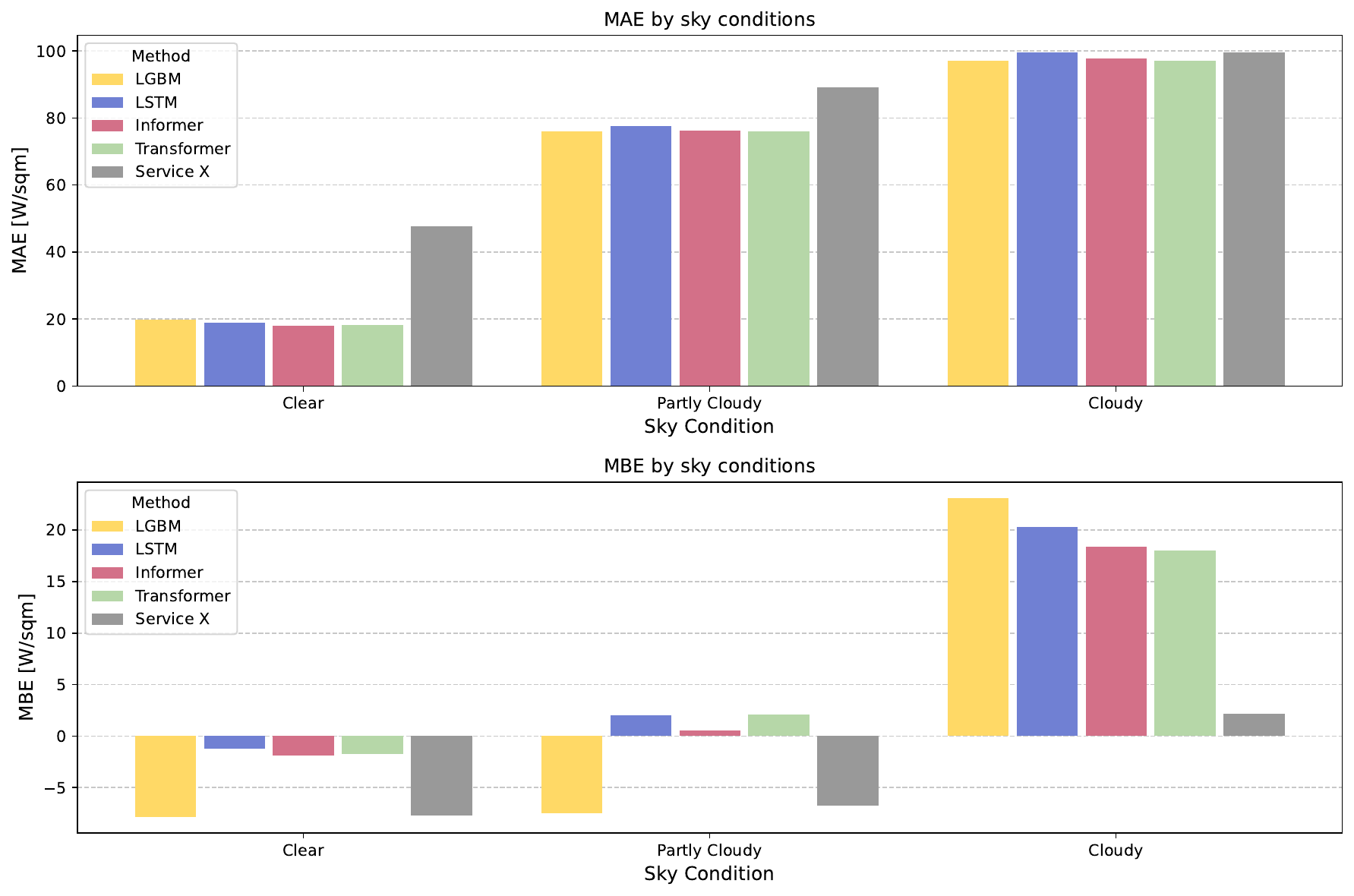}
\caption{Satellite-derived GHI estimation from benchmarked models: \emph{(Top.)} Validated MAE. \emph{(Bottom.)} MBE }
\label{fig:model_mae_mbe_byskycondition}
\end{figure}

\Cref{fig:model_mae_mbe_byskycondition} and \Cref{tab:model_error_matrix} show a comparative performance of all GHI estimation models, including LGBM, LSTM, Informer, Transformer, and the service X. All the proposed benchmarked models provide similar overall MAEs in the range of 78-81 $\wm$. For the MAE, the Informer, Transformer, and service X have the lowest error in clear-sky, partly cloudy, and cloudy conditions, respectively. For the MBE, LGBM and the service X have a relatively high negative bias (underestimate) in clear and partly cloudy condition. Meanwhile, LSTM, Informer, and Transformer tend to overestimate---high positive bias in the cloudy condition, while the service X has the lowest MBE in this case, as shown in \Cref{fig:model_mae_mbe_byskycondition}~\emph{(Bottom.)} The evaluation with the RMSE metric is reported in \Cref{tab:model_error_matrix}. All the proposed models perform almost equally if MAE is considered. For specific sky conditions, our proposed models generally perform well in clear-sky and partly cloudy scenarios. However, service X demonstrates superior performance under cloudy conditions.

\newcolumntype{k}{>{\columncolor[HTML]{FAE7B5}} C{1.5cm}}
\newcolumntype{z}{>{\columncolor[HTML]{ACE5EE}} C{1.5cm}}
\newcolumntype{t}{>{\columncolor[HTML]{F4C2C2}} C{1.5cm}}
\newcolumntype{o}{>{\columncolor[HTML]{9AB973}} C{2cm}}
\newcolumntype{s}{>{\columncolor[HTML]{DCDCDC}} C{1.5cm}}

\begin{table}[ht]
\tablefontsize
\centering
\caption{Model comparison of performance metrics: MAE, RMSE, and MBE ($\wm$) under three sky conditions.}
    \begin{tabular}{|lkztos|} \hline
    \bf Model & \bf LGBM & \bf LSTM & \bf Informer & \bf Transformer & \bf Service X \\ \hline
\multicolumn{6}{|l|}{\bf MAE ($\wm$)} \\ \hline
\bf Overall MAE & 78.58 & 80.41 & 78.97 & \textbf{78.52} & 80.80 \\
Clear sky & 19.73 & 18.92 & \textbf{17.99} & 18.11 & 27.00 \\
Partly cloudy & 75.93 & 77.62 & 76.30 & \textbf{75.90} & 80.16 \\
Cloudy & 97.01 & 99.68 & 97.81 & 97.07 & \textbf{91.99} \\ \hline
\multicolumn{6}{|l|}{\bf RMSE ($\wm$)} \\ \hline
\bf Overall RMSE & \textbf{118.97} & 125.86 & 124.54 & 123.82 & 121.99 \\
Clear sky & 36.70 & 32.79 & \textbf{31.54} & 31.94 & 39.52 \\
Partly cloudy & \textbf{115.13} & 122.19 & 121.21 & 120.58 & 120.58  \\
Cloudy & 138.53 & 145.74 & 143.45 & 142.42 & \textbf{135.01} \\ \hline
\multicolumn{6}{|l|}{\bf MBE ($\wm$)} \\ \hline
\bf  Overall MBE & \textbf{-0.47} & 6.11 & 4.55 & 5.59 & -4.21 \\
Clear sky & -7.86 & \textbf{-1.24} & -1.86 & -1.71 & -5.62 \\
Partly cloudy & -7.48 & 2.04 & \textbf{0.53} & 2.09 & -6.17 \\
Cloudy & 23.07 & 20.29 & 18.41 & 17.97 & \textbf{2.26} \\ \hline
    \end{tabular}
\label{tab:model_error_matrix}
\end{table}

\paragraph{Comparison with literature} Our GHI estimation at a half-hourly resolution provides the RMSE values (31.54 - 145.74 $\wm$) that aligns with previous studies in equatorial regions, \ie, ~\cite{Valor2023} reported hourly GHI RMSEs of 115-180$\wm$ across six locations (SAP, TAT, FUA, ISH, DWN, and ASP) using Himawari-8 data; and, ~\cite{Qin2021} reported an instantaneous GHI RMSE of 118.3 $\wm$ at the DWN station, also using Himawari-8. Our results are in a similar trend to~\cite{Kashyap2018}, providing half-hourly GHI RMSEs of 41.7-167 $\wm$ in South Asia (India).~\cite{BrightSolCast2019} reported an average hourly RMSE of 99.5 W/m² in equatorial regions for zenith angles between 0 and 85 degrees (\approxtext~05:56-18:37), exceeding our target forecasting hours (7:00-17:00). Note that including GHI estimates from early morning and late evening (nearly night-time) in the analysis could potentially underestimate the MAE and RMSE.

\paragraph{Impact of selected features} We explore possibly different characteristics of models when some features are not included in the model. NWP forecasts (both irradiance and temperature) are the input that is likely to be costly in operational setting when inferring a solar map of Thailand since the forecast values must be obtained a high spatial resolution. The coordinates of station locations are also considered to be removed, as this information is already inherited in the clear-sky model. \Cref{tab:mae_feature_removed} shows that dropping NWP input affects the performance slightly for all models and all-sky conditions, while the location coordinate still has some significance as the MAE is increased around 2 $\wm$. The Informer has the least MAE among all models when all two features are removed. Thus, we will focus on the performance of the informer.

\begin{table}[ht]
\tablefontsize
\centering
\caption{MAE comparison $(\wm)$ when some features are removed.}
\begin{tabular}{|l|c|c|c|c|} \hline 
\bf Model & \bf LGBM & \bf LSTM & \bf Informer & \bf Transformer  \\ \hline
\rowcolor{buff} \multicolumn{5}{|l|}{\bf Overall} \\ \hline
all features & \textbf{78.58} & 80.41 & 78.97 & 78.52 \\
no NWP & 78.98 & 80.77 & \textbf{78.67} & 79.06 \\
no NWP + Lat/Long & 82.35 & 82.05 & \textbf{80.85} & 80.98 \\ \hline
\rowcolor{palecornflowerblue} \multicolumn{5}{|l|}{\bf Clear sky} \\ \hline
all features & 19.73 & 18.92 & \textbf{17.99} & 18.11 \\
no NWP & 20.41 & 20.11 & \textbf{17.90} & 18.75 \\
no NWP + Lat/Long & 24.38 & 22.21 & \textbf{21.15} & 21.93 \\ \hline
\rowcolor{lightgray} \multicolumn{5}{|l|}{\bf Partly cloudy} \\ \hline
all features & 75.93 & 77.62 & 76.30 & \textbf{75.90} \\
no NWP & 76.27 & 77.86 & \textbf{75.74} & 76.38 \\
no NWP + Lat/Long & 79.97 & 79.50 & \textbf{78.29} & 78.36 \\\hline
\rowcolor{pink-orange} \multicolumn{5}{|l|}{\bf Cloudy} \\ \hline
all features & \textbf{97.01} & 99.68 & 97.81 & 97.07 \\
no NWP & \textbf{97.56} & 100.31 & 98.30 & 97.80 \\
no NWP + Lat/Long & 99.76 & 100.51 & \textbf{99.12} & 99.31 \\ \hline
\end{tabular}
\label{tab:mae_feature_removed}
\end{table}
\medskip

\begin{figure}
\centering
\begin{subfigure}{0.9\linewidth}
\centering
\includegraphics[width=\linewidth]{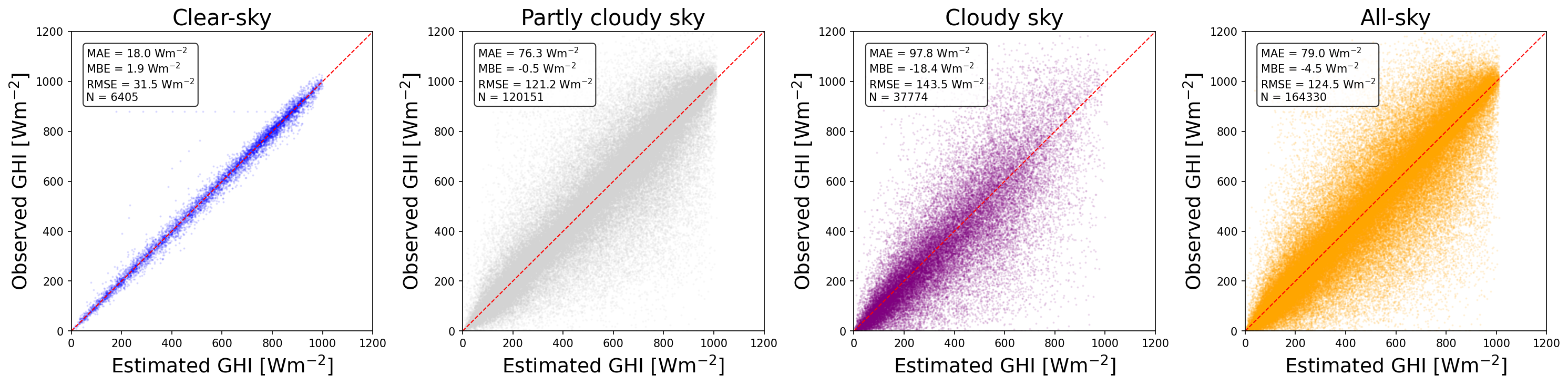}
\caption{Scatter plot between the estimated and actual irradiance.}
\end{subfigure} \\
\begin{subfigure}{1\linewidth}
\centering
\includegraphics[width=0.9\linewidth]{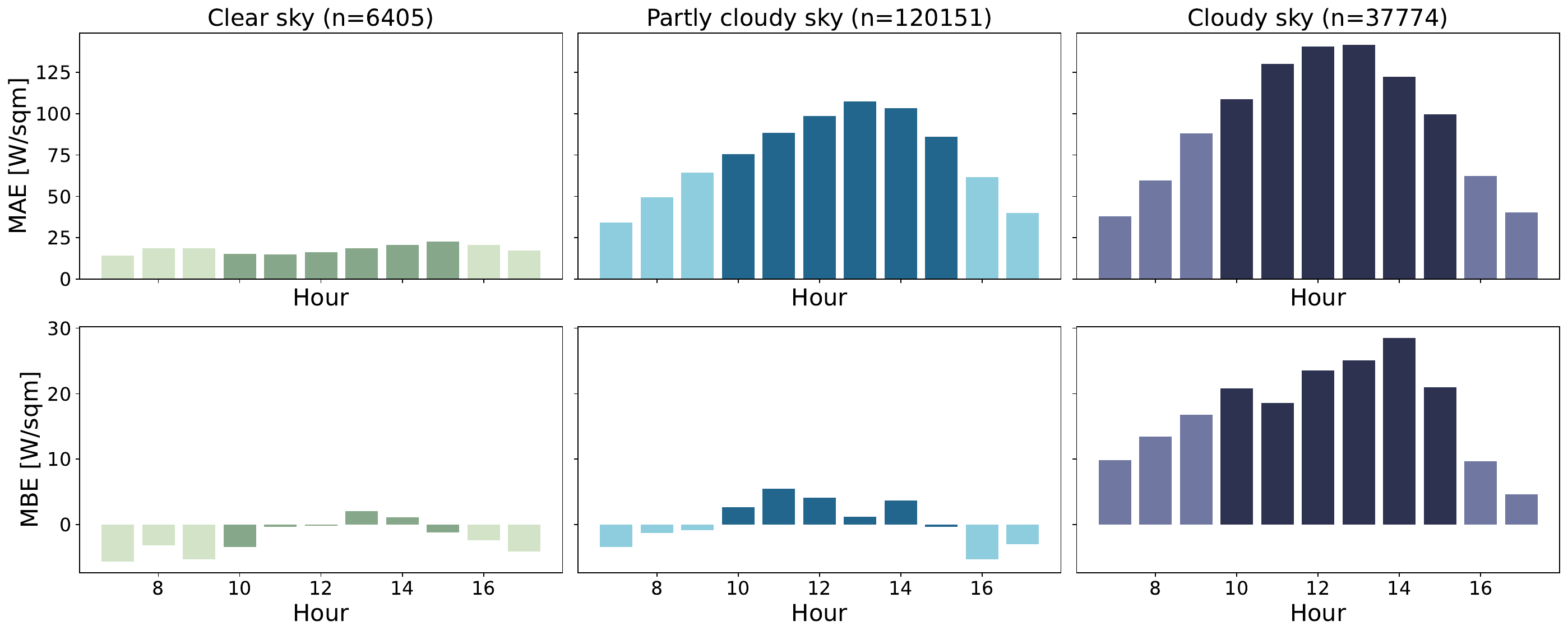}
\caption{MAE and MBE across three sky conditions and hours of day.}
\end{subfigure} \\
\begin{subfigure}{1\linewidth}
\centering
\includegraphics[width=0.9\linewidth]{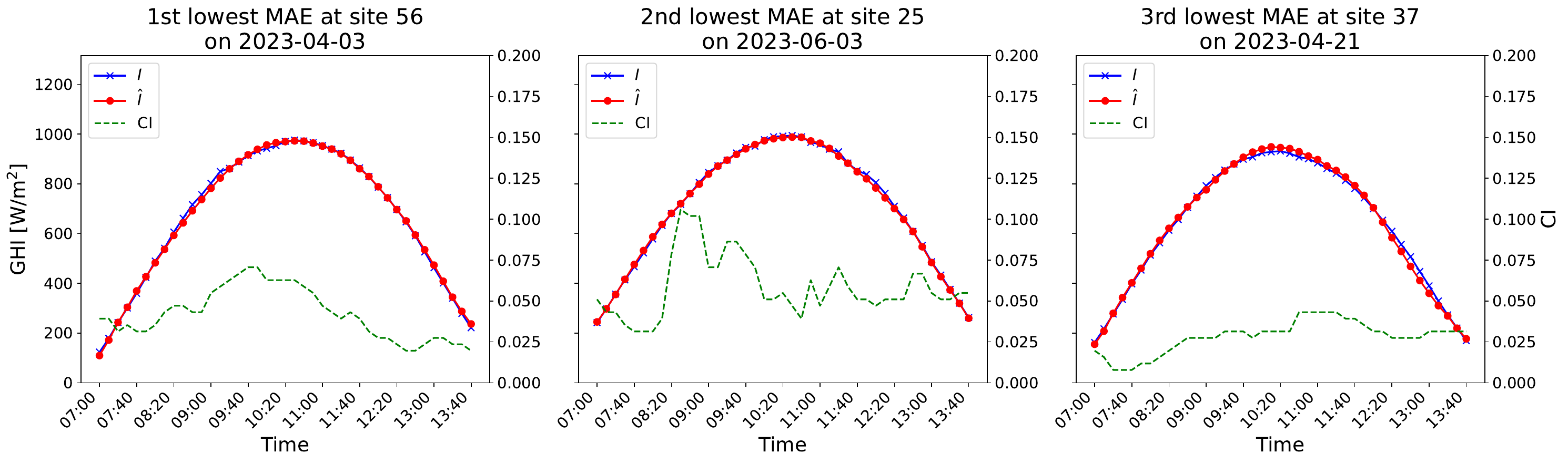}
\caption{Three \textbf{lowest MAE}.}
\end{subfigure} \\
\begin{subfigure}{1\linewidth}
\centering
\includegraphics[width=0.9\linewidth]{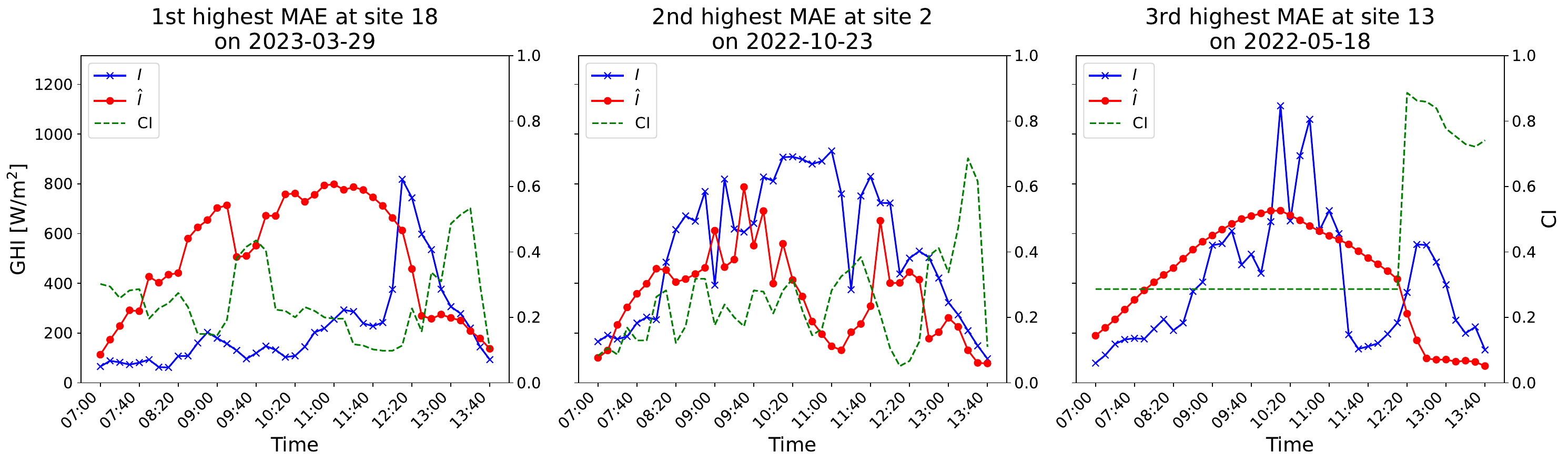}
\caption{Three \textbf{highest MAE}.}
\end{subfigure}
\caption{Solar irradiance estimation by Informer: (a) Discrepancy against the actual irradiance and (b) MAE and MBE across three sky conditions and hours of the day. The samples of the selected time series with (d) the top three lowest and (e) the top three highest daily average MAE.}
\label{fig:bestmodel}
\end{figure}

The solar irradiance estimation by Informer is provided in \Cref{fig:bestmodel}. The discrepancies between the estimates and ground truth appear to be larger when the sky is cloudy, as shown in \Cref{fig:bestmodel} (a). Meanwhile, the MAE evaluated for each hour of the day indicates that the model usually suffers during the daytime (10:00-15:00) under the presence of cloud, as shown in \Cref{fig:bestmodel} (b). The MAE could go up to 140 $\wm$ in this period, and the positive MBE explains that the model mostly overestimates in this period. \Cref{fig:bestmodel} (c) and (d) show the selected time series with the top three lowest and highest daily average MAE, respectively. The time series with the lowest error often co-occurs with the clear sky condition, where the cloud index is relatively low. The time series with the highest error typically happens with cloud fluctuation during the day. Thus, the model functioned as expected; if CI increases, the estimate $\hat{I}$ decreases. A high estimation error, for example, the middle of the plot (d), could come from approximation in the data processing step. It could be the case that CI was missing before 14:00, and the displayed values were interpolated (as a constant of 0.3 from backward imputation). 

\subsection{Solar energy in Thailand}
\label{sec:solarenergy}
\begin{figure}[h!]
\centering
\begin{subfigure}[b]{0.3\textwidth}
\centering
    \includegraphics[width=\linewidth]{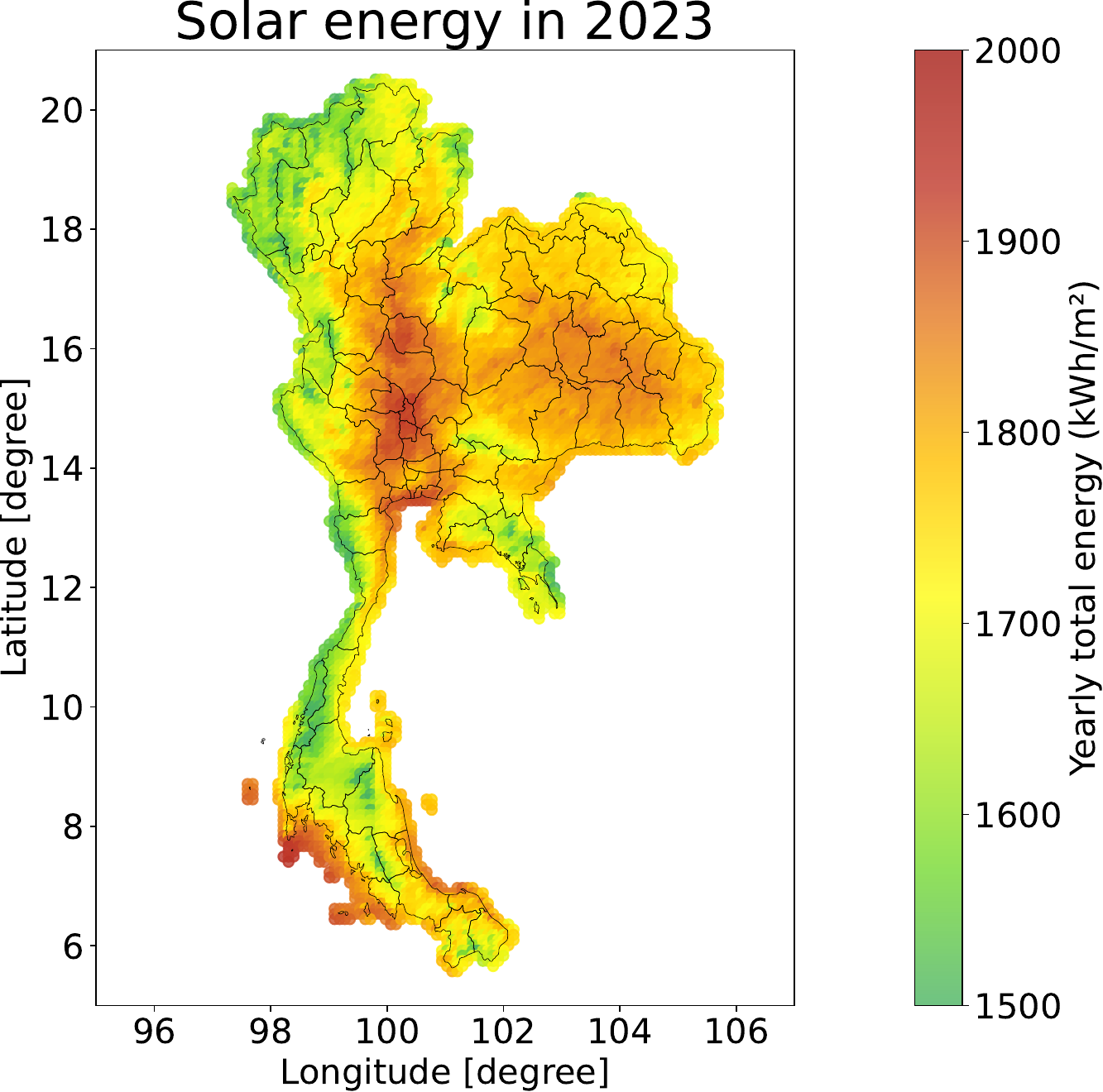}
\caption{\textbf{Yearly}}
\label{fig:map_yearly_energy_2023}
\end{subfigure}
\begin{subfigure}[b]{0.65\textwidth}
\centering
\includegraphics[width=\linewidth]{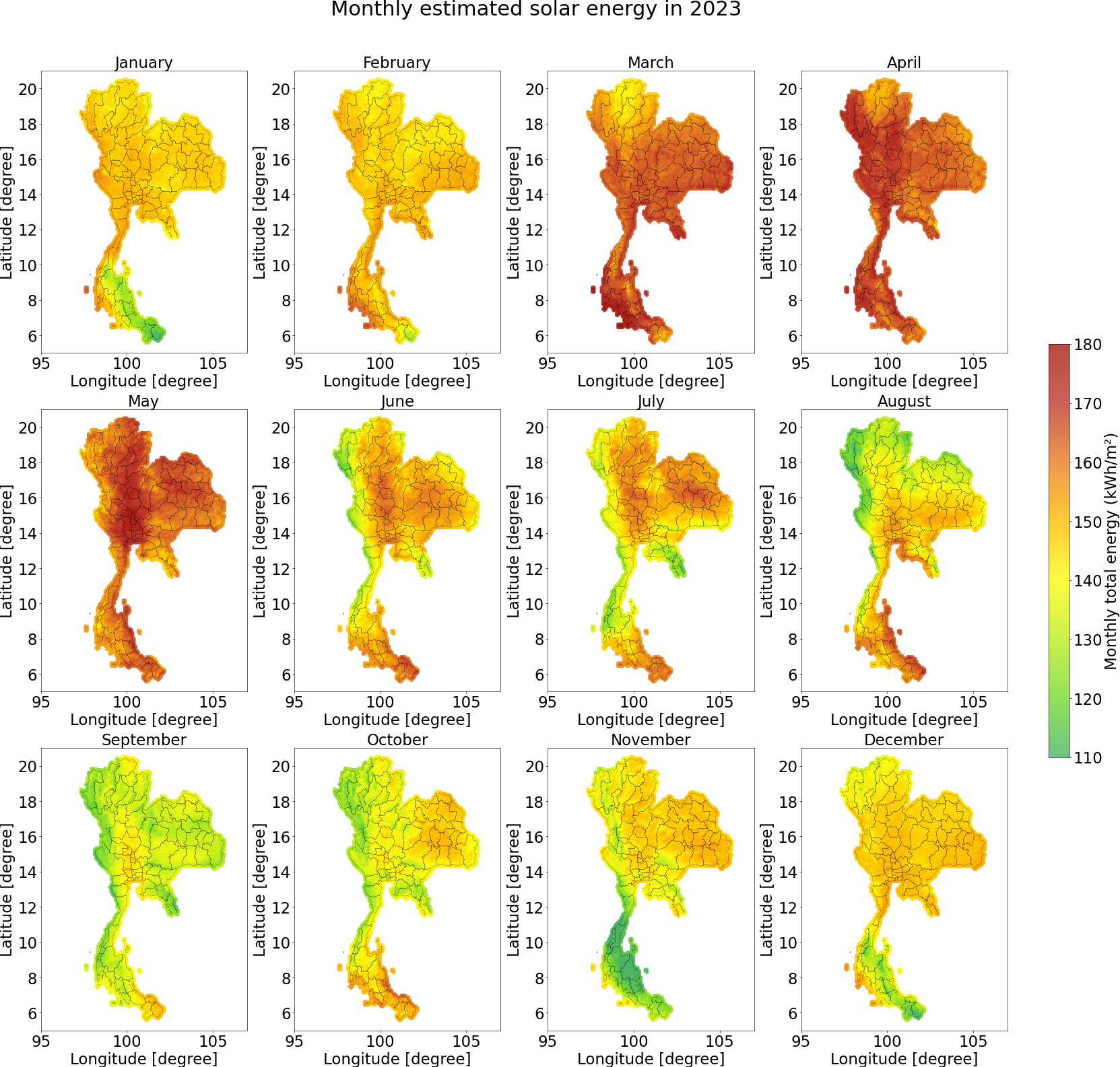}
\caption{\textbf{Monthly}}
\label{fig:map_monthly_energy}
\end{subfigure} \\
\begin{subfigure}[b]{1\textwidth}
\centering
\includegraphics[height = 0.5\textwidth]{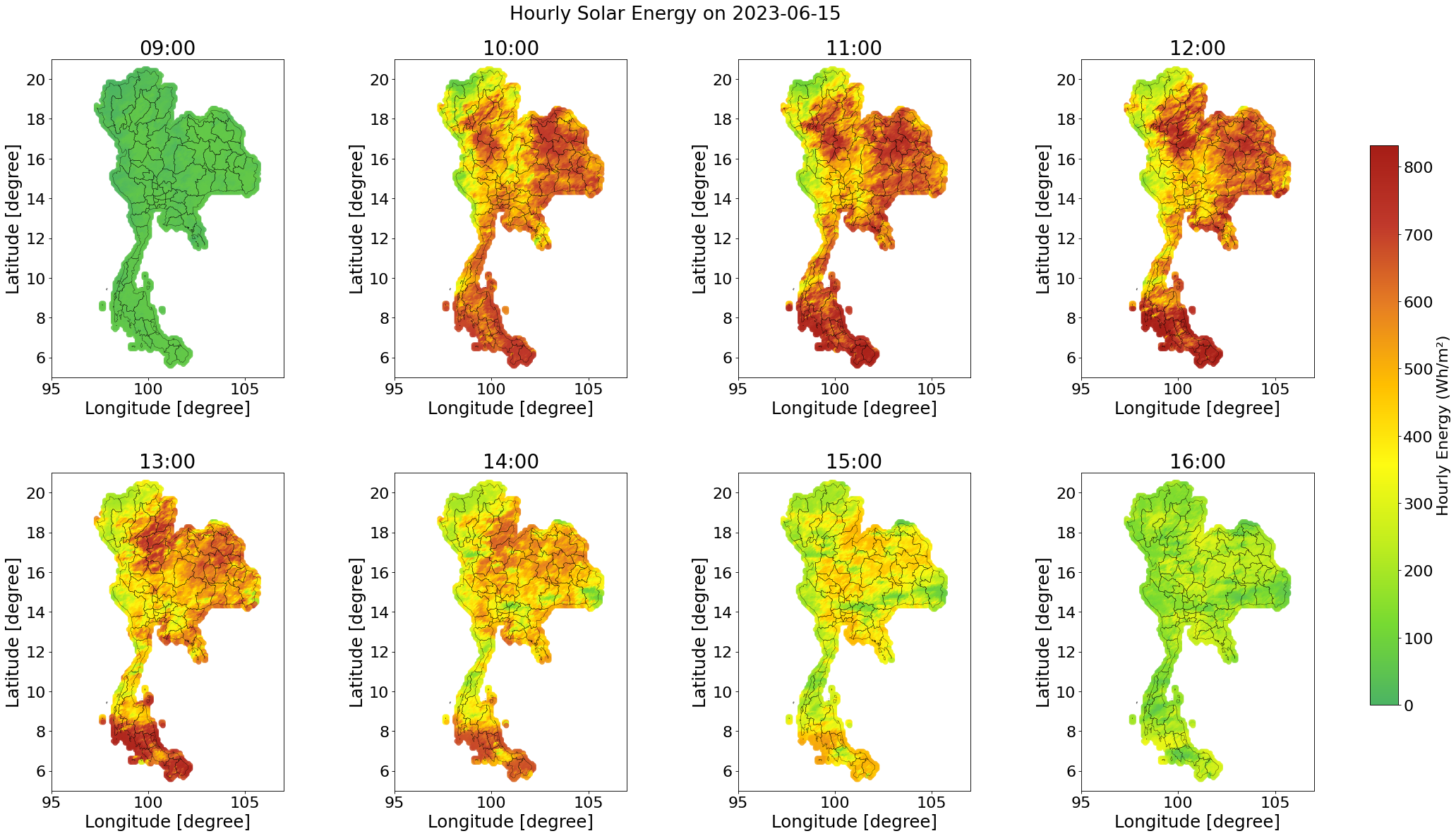}
\caption{\textbf{Hourly} on 15/06/2023.}
\label{fig:map_hourly_energy}
\end{subfigure}
\caption{Estimated total solar energy in 2023, Thailand, by the Informer model.}
\label{fig:estsolarmap}
\end{figure}

Estimated GHI across Thailand by Informer are cumulatively summed up in yearly, monthly, and hourly time scales of 2023 to infer a solar energy map of Thailand shown in \Cref{fig:estsolarmap}.

We compare our results with previous studies that visualized solar energy maps in nearby countries. Ministry of Energy, Thailand, currently has developed a yearly solar energy map  \cite{dedesolarmap, Janjai2013} using long-term data from GMS-5, GOES-9, MTSAT, and Himawari satellites. The methodology relied on physical and semi-empirical models presented in \cite{Janjai2013} that require reflectivity and absorption by water vapor, ozone, and particles to infer global solar radiation. \cite{Janjai2011} presented an application for investigating potential areas in Thailand for producing electricity from solar power. The method used GMS5 satellite data from 1995-2002, where the variables include reflectivity and absorption by water vapor, which was computed by ambient temperature and relative humidity. The map presented the year sum of DNI. \cite{Rumbayan2012} visualized monthly solar energy in Indonesia using estimates from an ANN model with weather inputs and geographical information. \cite{Siala2016} used the historical MERRA solar radiation dataset to analyze potential solar power areas in Southeast Asia. Eastern Thailand, Java, and central Myanmar were the best locations for solar power use.

\Cref{fig:map_yearly_energy_2023} shows total yearly energy where effective areas yielding high solar energy lie in central and northeastern regions; the result aligns with that of the solar radiation map provided by DeDe (\url{https://gis.dede.go.th/upload/20230302_135735.pdf}); see the Appendix. \Cref{fig:map_monthly_energy}  
shows the monthly estimated energy, with most of Thailand experiencing the highest levels of 170-180 $\kwhm$ in March, April, and May. This aligns with the monthly average of daily global radiation in 2013~\cite{Janjai2013}, which peaks at approximately 183 $\kwhm$ or 22 $\text{MJ}/\text{m}^2-\text{day}$ during the same months. However, due to missing data (52\% in July and 67\% in November), the monthly GHI estimates for these months are extrapolated, resulting in a solar map that differs from~\cite{Janjai2013}.     

\Cref{fig:map_hourly_energy} illustrates hourly-updated GHI estimates on June 15, 2023. Our GHI estimation model can provide solar energy estimation at the finer temporal resolution, compared to~\cite{Janjai2013}, which was aimed at analyzing a monthly resolution. Our model prioritizes near real-time updates, which offers a significant advantage for operational settings in energy applications where real-time updates are crucial.

\subsection{Deployment and runtime}

Our GHI models are deployed to provide the solar map at two spatial resolutions: zoom level 12 (6,843 grids, approximately $\approx 9.8 \times 9.8~\sqkm$ per grid) and zoom level 14 (93,061 grids, approximately $\approx 2.5 \times 2.5~\sqkm$ per grid). To efficiently estimate solar irradiance for all grid points simultaneously, we set the batch size equal to the total number of grids in each map. The runtime performance of LSTM, Informer, and Transformer models on an RTX4060 GPU for this process is presented in \Cref{tab:runtime}. From \Cref{sec:LGBM}, when directly including latitude and longitude as features for LGBM, the generated solar map displays some abrupt changes in estimated GHI at some region boundary. These unrealistic results can be fixed when removing these two features, with the cost of having 2-3 $\wm$ MAE increased.
For this reason, LGBM results are excluded from the runtime report. All models show acceptable performance ($\leq 2$ secs. for the largest batch size), with LSTM being the fastest. The Transformer can be an alternative to the Informer because of their similar performance. 
 
\begin{table}[ht]
\tablefontsize
\centering
\caption{Average performance per run: inference runtime, overall runtime (inference + data loading), and peak memory.}
\begin{tabular}{cl|c|c|c}  \hline
                        & \bf DL models   & \bf Inference (ms) & \bf Overall (ms) & \bf Peak memory (MB)  \\ \hline
\multirow{3}{2.25cm}{\centering \textit{Zoom level 12} (Batch~size~=~6,843)}    & LSTM        & 22.35           & 123.89         & 417.86   \\
                                                                    & Informer    & 72.29           & 151.47         & 581.50  \\
                                                                    & Transformer & 67.83           & 151.70         & 583.69 \\   \hline
\multirow{3}{2.25cm}{\centering \textit{Zoom level 14} (Batch~size~=~93,601)}   & LSTM        & 307.43          & 1,757.57       & 5,451.18   \\
                                                                    & Informer    & 561.87          & 1,999.29       & 7,754.21\\
                                                                    & Transformer & 500.84          & 1,946.88       & 7,783.98 \\   \hline
\end{tabular}
\label{tab:runtime}
\end{table} 

\section{Conclusions}
This research presented a methodology of a satellite-derived GHI estimation across Thailand with a spatial resolution of $2.4 \times 2.4 \; \sqkm$, totaling 93,061 grids with 30-minute updates using cloud indices extracted from Himawari-8 satellite imageries, along with the output of Ineichen and Perez's clear-sky irradiance model as main features of our GHI estimation models.  The study explored recently effective machine learning models, including LightGBM (LGBM), LSTM, Informer, and Transformer, to capture complex cloud effects on ground irradiance attenuation.  To optimize computational efficiency, input data is processed as time series, with necessary modifications to the Informer and Transformer. Benchmarking against a commercial service from 53 ground stations demonstrates an improved performance, with lower MAE and RMSE values. Our GHI estimates, at a half-hourly resolution and with RMSE values ranging from 31.54 to 142.42 W/m², align with previous studies~ \cite{Kashyap2018, BrightSolCast2019, Qin2021, Valor2023}. We provide real-time solar energy maps at hourly, monthly, and yearly resolutions, and assess computational efficiency at different spatial scales. Our monthly results on March, April, and May are aligned with the monthly average of daily global irradiance in 2013 during the same months~\cite{Janjai2013}. Additionally, we confirmed the feasibility of using these models for real-time GHI estimation at the required temporal resolution.

\section{Acknowledgment}
The authors thank Parinthorn Manomaisaowapak for developing the GHI estimation model (version-2), Justin Jiraratsakul and NDR solution (Thailand) Co., Ltd. for implementing the solar map platform as well as Tanan Boonyasirikul for the initial experiments on the DL models.  We appreciate the discussions with the Division of Solar Energy Development, Department of Alternative Energy Development and Efficiency (DeDe), Ministry of Energy, which provided supplementary solar data and a historical yearly solar map. This work was co-funded by the Chula Engineering Research grant 2023 and partially supported by Grants for Development of New Faculty Staff, Ratchadaphiseksomphot Endowment Fund. 


\bibliographystyle{alpha} 
\bibliography{ref_solarmapCUEE}

\newpage
\section*{Appendix: Supplementary results}
\addcontentsline{toc}{section}{Appendix}

\begin{figure}[ht]
\centering
\includegraphics[width=0.6\linewidth]{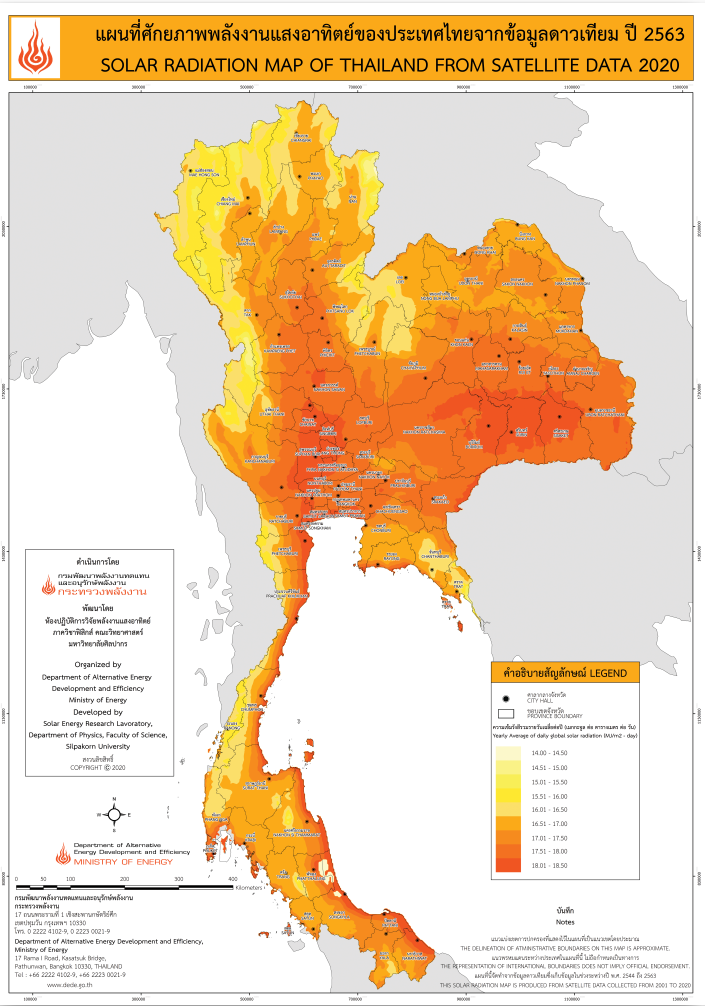}
\caption{Solar radiation map of 2023, provided by Ministry of Energy (Source: \url{https://gis.dede.go.th/upload/20230302_135735.pdf}). The map shows yearly average of daily global radiation ($\mathrm{MJ/m}^2 \cdot \mathrm{day}$).}
\label{fig:dede_solarmap}
\end{figure}

\Cref{tab:solarservice} lists available services that provide satellite-derived solar irradiance. \Cref{tab:performance53sites} shows the detailed estimation performance of the Informer model for each ground station. \Cref{fig:dede_solarmap} gives a comparative estimated solar energy map from Ministry of Energy that was discussed in \Cref{sec:solarenergy}. 

\begin{table}[ht]
\tablefontsize
\caption{Satellite-derived solar irradiance services.}
\begin{tabular}{|p{0.25\textwidth} p{0.38\textwidth}  p{0.35\textwidth} |} \hline
\bf Service & \bf Method & \bf  Web page \\ \hline
\rowcolor{buff} \multicolumn{3}{|l|}{\bf Global} \\  
SolarAnywhere & SUNY model & https://www.solaranywhere.com/ \\
NSRDB &  Hybrid SUNY/METSTAT models & https://nsrdb.nrel.gov/ \\
SolarGIS & 
SolarGIS satellite-to-irradiance model & https://solargis.com/ \\
SolCast & SolCast cloud model & https://solcast.com/ \\ 
GEWEX-SRB & Radiative transfer algorithm   & https://science.larc.nasa.gov/gewex-srb/ \\ 
Meteonorm 8 & Stochastic generation of global radiation & https://meteonorm.com/ \\
Vaisala & Vaisala algorithm (REST2/Krasen models)  & https://www.vaisala.com/en \\ 
CM SAF (SARAH 3) & MAGICSOL & https://wui.cmsaf.eu/ \\
\rowcolor{buff} \multicolumn{3}{|l|}{\bf Europe, Africa, and Parts of Asia and South America} \\
Soda-Pro & HelioClim-3~(Heliosat-2+McClear) &  https://www.soda-pro.com/ \\

\rowcolor{buff} \multicolumn{3}{|l|}{\bf Europe, Africa, and Parts of Asia, South America and Australia } \\  
PVGIS & SARAH and NSRDB & https://pvgis.com/ \\
CAMS Radiation service & McClear, McCloud models and Heliosat-4 & https://atmosphere.copernicus.eu/\\ 
SOLEMI & Heliosat method  & https://www.wdc.dlr.de/data-products/solar-radiation/solemi-data-base \\
\rowcolor{buff} \multicolumn{3}{|l|}{\bf Australia} \\ 
Australian Bureau of Meteorology & Physical model  & http://www.bom.gov.au/climate/ \\ \hline
\end{tabular}
\label{tab:solarservice}
\end{table}

\newpage
\begin{footnotesize}
\begin{longtable}{|c|l|c|c|c|c|c|c|c|}
    \caption{GHI estimation performance metrics of the Informer model split by ground stations. MAE, RMSE, MBE, $\bar{I}$ are in $\wm$ and NMAE, NRMSE, NMBE are in \%.} \\ \hline
    \footnotesize
\bf ID & \bf Site name & \bf MAE & \bf RMSE & \bf MBE & \bf NMAE & \bf NRMSE & \bf NMBE & \bf $\bar{I}$  \\ \hline
\endfirsthead
\multicolumn{9}{c}%
{{\bfseries \tablename\ \thetable{} -- continued from previous page}} \\
\hline 
\bf ID & \bf Site Name & \bf MAE & \bf RMSE & \bf MBE & \bf NMAE & \bf NRMSE & \bf NMBE & \bf $\bar{I}$  \\ \hline
\endhead

\hline \multicolumn{9}{|r|}{{Continued on next page}} \\ \hline
\endfoot

\hline \hline
\endlastfoot

\rowcolor{buff} \multicolumn{9}{|l|}{\bf Central} \\ \hline   
 1 & Pathum Thani 1 & 71.71 & 109.81 & -0.21 & 14.96 & 22.91 & -0.04 & 479.32 \\ \hline
 2 & Nakhon Pathom 1 & 115.36 & 174.97 & -7.18 & 23.35 & 35.42 & -1.45 & 494.03 \\ \hline
 3 & Samut Prakan 1 & 93.56 & 149.26 & 15.90 & 19.47 & 31.07 & 3.31 & 480.47 \\ \hline
 4 & Samut Prakan 2 & 73.55 & 115.33 & -0.48 & 14.33 & 22.48 & -0.09 & 513.12 \\ \hline
 5 & Chachoengsao & 80.48 & 120.07 & 7.84 & 15.95 & 23.80 & 1.55 & 504.49 \\ \hline
 6 & Nonthaburi 1 & 107.18 & 168.64 & 3.77 & 21.93 & 34.51 & 0.77 & 488.70 \\ \hline
 7 & Nakhon Sawan & 79.97 & 120.04 & -6.54 & 15.82 & 23.74 & -1.29 & 505.65 \\ \hline
 8 & Phra Nakhon Si Ayutthaya & 77.72 & 123.52 & -2.78 & 15.11 & 24.01 & -0.54 & 514.37 \\ \hline
 10 & Nonthaburi 2 & 79.60 & 123.01 & 3.65 & 15.76 & 24.36 & 0.72 & 504.91 \\ \hline
 11 & Samut Prakan 3 & 76.21 & 114.40 & 6.72 & 14.94 & 22.42 & 1.32 & 510.20 \\ \hline
 33 & Pathum Thani 2 & 80.26 & 122.39 & 12.81 & 16.10 & 24.55 & 2.57 & 498.46 \\ \hline
 35 & Nakhon Pathom 2 & 73.42 & 115.35 & -0.49 & 14.40 & 22.62 & -0.10 & 510.01 \\ \hline
 36 & Pathum Thani 3 & 73.03 & 112.39 & -0.02 & 15.14 & 23.30 & -0.00 & 482.43 \\ \hline
 42 & Pathum Thani 4 & 79.13 & 125.13 & 7.37 & 15.22 & 24.06 & 1.42 & 520.07 \\ \hline
 46 & Samut Prakan 4 & 67.04 & 103.97 & -6.36 & 13.03 & 20.21 & -1.24 & 514.39 \\ \hline
 47 & Bangkok 1 & 82.89 & 122.37 & 16.66 & 18.38 & 27.13 & 3.69 & 451.05 \\ \hline
 48 & Bangkok 2 & 73.99 & 111.52 & 16.96 & 15.81 & 23.82 & 3.62 & 468.12 \\ \hline
 56 & Samut Sakhon & 61.68 & 103.12 & 7.82 & 11.65 & 19.48 & 1.48 & 529.49 \\ \hline
\rowcolor{pink-orange} \multicolumn{9}{|l|}{\bf North} \\  \hline 
9 & Phrae & 73.68 & 117.37 & -9.74 & 14.75 & 23.50 & -1.95 & 499.46 \\ \hline
15 & Lampang & 78.89 & 124.89 & 25.24 & 16.01 & 25.34 & 5.12 & 492.87 \\ \hline
19 & Sukhothai & 68.24 & 111.59 & -1.55 & 12.99 & 21.24 & -0.30 & 525.40 \\ \hline
20 & Lamphun & 70.01 & 111.42 & 5.81 & 13.87 & 22.08 & 1.15 & 504.70 \\ \hline
27 & Chiang Mai 1 & 76.71 & 123.49 & -0.45 & 13.90 & 22.38 & -0.08 & 551.72 \\ \hline
28 & Chiang Mai 2 & 77.01 & 129.11 & -1.80 & 15.07 & 25.26 & -0.35 & 511.03 \\ \hline
38 & Phetchabun 1 & 77.26 & 125.28 & -1.37 & 15.02 & 24.35 & -0.27 & 514.42 \\ \hline
43 & Phetchabun 2 & 68.89 & 111.56 & 7.04 & 13.38 & 21.66 & 1.37 & 515.07 \\ \hline
\rowcolor{palecornflowerblue} \multicolumn{9}{|l|}{\bf NorthEast} \\ \hline 
12 & Yasothon & 70.24 & 111.73 & 5.66 & 13.41 & 21.33 & 1.08 & 523.92 \\ \hline
13 & Ubon Ratchathani & 112.93 & 154.23 & 1.33 & 23.80 & 32.50 & 0.28 & 474.52 \\ \hline
17 & Buriram & 80.50 & 122.44 & 6.77 & 16.20 & 24.64 & 1.36 & 496.92 \\ \hline
18 & Kalasin & 122.08 & 186.83 & 16.75 & 24.38 & 37.30 & 3.34 & 500.85 \\ \hline
21 & Udon Thani & 98.28 & 150.02 & 17.39 & 19.90 & 30.38 & 3.52 & 493.84 \\ \hline
22 & Khon Kaen & 71.75 & 116.13 & 7.55 & 13.43 & 21.74 & 1.41 & 534.24 \\ \hline
25 & Chaiyaphum & 62.80 & 100.47 & -7.24 & 11.83 & 18.92 & -1.36 & 531.02 \\ \hline
26 & Nakhon Ratchasima 1 & 72.83 & 119.11 & 5.90 & 14.16 & 23.16 & 1.15 & 514.34 \\ \hline
30 & Nakhon Ratchasima 2 & 79.85 & 128.24 & 15.90 & 15.59 & 25.03 & 3.10 & 512.33 \\ \hline
31 & Sakon Nakhon 1 & 67.72 & 112.70 & -0.10 & 13.67 & 22.75 & -0.02 & 495.27 \\ \hline
39 & Sakon Nakhon 2 & 88.73 & 137.80 & 18.89 & 17.21 & 26.73 & 3.66 & 515.52 \\ \hline
41 & Roi Et & 74.98 & 121.24 & 14.23 & 14.37 & 23.24 & 2.73 & 521.80 \\ \hline
\rowcolor{lightgray} \multicolumn{9}{|l|}{\bf East} \\  \hline 
23 & Chonburi 1 & 71.68 & 117.43 & -11.20 & 14.24 & 23.33 & -2.22 & 503.30 \\ \hline
24 & Chanthaburi & 86.62 & 136.04 & -8.11 & 18.83 & 29.57 & -1.76 & 459.98 \\ \hline
40 & Chonburi 2 & 74.47 & 120.37 & 0.11 & 15.55 & 25.14 & 0.02 & 478.84 \\ \hline
44 & Sa Kaeo & 79.83 & 121.06 & -7.05 & 15.74 & 23.87 & -1.39 & 507.25 \\ \hline
49 & Chonburi 3 & 71.88 & 110.06 & 16.03 & 15.11 & 23.14 & 3.37 & 475.70 \\ \hline
50 & Chonburi 4 & 71.60 & 117.32 & 2.53 & 14.41 & 23.62 & 0.51 & 496.79 \\ \hline
51 & Chonburi 5 & 66.96 & 106.10 & -0.84 & 13.56 & 21.48 & -0.17 & 493.91 \\ \hline
52 & Chonburi 6 & 76.58 & 113.42 & 14.25 & 17.14 & 25.39 & 3.19 & 446.67 \\ \hline
53 & Chonburi 7 & 81.64 & 121.07 & -1.96 & 16.82 & 24.94 & -0.40 & 485.35 \\ \hline
55 & Chonburi 8 & 70.57 & 106.89 & 8.06 & 14.52 & 21.99 & 1.66 & 486.15 \\ \hline
\rowcolor{lightcoral} \multicolumn{9}{|l|}{\bf West} \\  \hline 
16 & Ratchaburi & 77.59 & 122.36 & 4.73 & 15.31 & 24.15 & 0.93 & 506.72 \\ \hline
37 & Phetchaburi & 65.51 & 105.03 & -2.07 & 12.46 & 19.98 & -0.39 & 525.78 \\ \hline
\rowcolor{mediumspringbud} \multicolumn{9}{|l|}{\bf South} \\ \hline 
14 & Surat Thani & 96.81 & 141.02 & 10.14 & 20.80 & 30.29 & 2.18 & 465.56 \\ \hline
29 & Krabi & 90.63 & 139.25 & 13.97 & 18.26 & 28.05 & 2.81 & 496.39 \\ \hline
34 & Nakhon Si Thammarat & 90.98 & 133.65 & -0.32 & 18.91 & 27.78 & -0.07 & 481.08 \\ \hline
\label{tab:performance53sites}
\end{longtable}
\end{footnotesize}

\newpage

We can convert the yearly average of daily global radiation ($\mathrm{MJ/m}^2 \cdot \mathrm{day}$) in \Cref{fig:dede_solarmap} by multiplying $0.2778 \cdot 365$ to be the unit of $\kwhm$ as to compared with  \Cref{fig:map_yearly_energy_2023} (total yearly energy). For example, the hottest and coldest spots in \Cref{fig:dede_solarmap} are 18.50 and 14.00 $\mathrm{MJ/m}^2 \cdot \mathrm{day}$, respectively, and they correspond to the total yearly energy of 1875.84 and 1419.56 $\kwhm$ (which can be compared to values in \Cref{fig:map_yearly_energy_2023}). As to compare with \Cref{fig:dede_solarmap}, our estimated map in \Cref{fig:map_yearly_energy_2023} has a similar trend of spatially relative energy while an actual value of yearly energy estimate at a specific location (except in northeastern) appears to be higher.

\end{document}